\documentclass[10pt,twocolumn]{article}

\usepackage{graphicx}
\usepackage{url}
\usepackage{mathtools}
\usepackage{amsmath,amssymb}
\usepackage[separate-uncertainty=true]{siunitx}
\usepackage{multirow}
\usepackage[percent]{overpic}
\usepackage{microtype}
\usepackage{hyperref}
\usepackage[dvipsnames,table,xcdraw]{xcolor}
\usepackage{makecell}
\usepackage{cite}
\usepackage{geometry}
\usepackage[makeroom]{cancel}
\usepackage{booktabs}
\usepackage[usestackEOL]{stackengine}
\usepackage{placeins}
\usepackage{tikz}
\usetikzlibrary{calc}
\usepackage{adjustbox}
\usepackage{balance}
\usepackage{dcolumn}
\usepackage{bm}

\usepackage{etoolbox}
\newtoggle{isPNAS}
\togglefalse{isPNAS} 

\definecolor{linkcolor}{rgb}{0.5,0.1,0.1}
\definecolor{urlcolor}{rgb}{0.1,0.1,0.5}
\definecolor{citecolor}{rgb}{0.1,0.5,0.1}

\hypersetup{
  colorlinks=true,
  linkcolor=linkcolor,
  urlcolor=urlcolor,
  citecolor=citecolor,
  pdfborder={0 0 0}
}

\graphicspath{{figs/}}

\newcommand{\delete}[1]{}

\renewcommand{\cancel}[1]{}

\title{
From rings to resonance: an inverse method links biophotonic structural color to inverse photonic glasses}

\author{
Florin Hemmann$^{1,2,3}$,
Matthias Saba$^{1,2}$,
Ullrich Steiner$^{1,2}$, \\
Mauro S. Ferreira$^{4,5}$,
Felipe A. Pinheiro$^{3}$\\[2ex]
\begin{tabular}{c}
{\small $^1$ Adolphe Merkle Institute, University of Fribourg, 1700 Fribourg, Switzerland}\\
{\small $^2$ NCCR Bio-Inspired Materials, University of Fribourg, 1700 Fribourg, Switzerland}\\
{\small $^3$ Instituto de Física, Universidade Federal do Rio de Janeiro, 21941-585 Rio de Janeiro, Brazil}\\
{\small $^4$ School of Physics, Trinity College Dublin, D02 PN40 Dublin 2, Ireland}\\
{\small $^5$ Centre for Research on Adaptive Nanostructures and Nanodevices (CRANN) and}\\
{\small \phantom{$^5$} Advanced Materials and Bioengineering Research (AMBER) Centre,}\\
{\small \phantom{$^5$} Trinity College Dublin, D02 CP49 Dublin 2, Ireland}
\end{tabular}
}

\begin{document}

\twocolumn[
\maketitle

\begin{abstract}
Structural color arises from the interaction of light with nanoscale structures and is widespread in nature. As structural complexity increases, the mechanisms governing coloration become progressively less understood. The optical response of periodic photonic crystals with a spatially periodic refractive index is well described by Bloch theory, whereas that of photonic glasses composed of randomly assembled uniform spheres is more subtle yet well studied. In contrast, disordered photonic networks found in many beetles are among the most complex natural photonic architectures, and the fundamental relationships between their structure and color remain unclear. Here, we use an inverse method to identify the structural features encoded in the reflectance spectrum. By comparing the spectrum of an unknown system with a database of simulated spectra from computer-generated photonic networks, we infer its structural properties. Applying this approach to both simulated networks and the biophotonic network responsible for the blue coloration of the weevil \textit{Pachyrhynchus congestus mirabilis}, we identify rings and pores as the key local scattering motifs. Their characteristic sizes govern the spectral position of the reflectance peak, whereas short-range disorder controls its width. The inverse method reveals clear spectral signatures of short-range order, whereas the influence of hyperuniformity and primitive similarity appears comparatively weak. This suggests that blue structural coloration is governed by local scattering mechanisms rather than photonic band-gap effects. We propose an analogy to an inverse photonic glass, in which pores and rings act as correlated local resonators. This perspective provides new design principles for bio-inspired structural-color materials.
\end{abstract}

\vspace{2\baselineskip}
]

\noindent Structural color originates from the interference of visible light in photonic nanostructures and is responsible for the vivid, durable coloration observed in many biological systems.\cite{vukusic2003, rothammer2021, djeghdi2022, vogler-neuling2023, bauernfeind2023, bauernfeind2024}. The reflectance spectrum and its angle dependence are determined by material properties, such as dielectric contrast, volume fraction $\phi$, length scale, spatial symmetries, and the geometry and topology of the unit cell or building blocks.
Structural color mechanisms can be classified by the material's spatial symmetries and dielectric contrast. 

Photonic crystals are systems with a spatially periodic refractive index \cite{joannopoulos2008}. Due to this periodicity, their eigenmodes are Bloch states. At high dielectric contrast, destructive interference can give rise to frequency domains without any non-evanescent eigenmode, so-called full photonic band gaps (PBGs). Incoming light at frequencies inside the PBG cannot couple to any electromagnetic mode, and therefore is fully reflected, and appears as structural color \cite{vukusic2003}. PBGs are favored by bicontinuous network morphologies, which permit the concentration of electromagnetic eigenstates near the band edges within one of the two constituent phases \cite{joannopoulos2008}. The PBG width is maximal when \cite{yeh2005, joannopoulos2008}
\begin{equation}
\phi n_\mathrm{m}=(1-\phi)n_\mathrm{b}
\label{eqn:volume_fraction_pbg}
\end{equation}
promotes optimal confinement of electromagnetic energy within a single phase,
\iftoggle{isPNAS}
{\Parasplit}{}
where
$n_\mathrm{m}$ and $n_\mathrm{b}$ are the refractive indices of the high-index and background media, respectively.
As the dielectric contrast decreases, the PBG closes, but a remnant reduced photonic density of states still leads to an enhanced reflection and structural color \cite{yin2012}.

Disordered photonic materials with high dielectric contrast can also exhibit PBGs, despite the absence of long-range order and a Bloch description \cite{edagawa2008}. 
Even without periodicity, hidden long-range order may arise, encoded in hyperuniform structural correlations, which favors PBGs \cite{florescu2009, torquato2018, siedentop2024}. 
For 2D systems, Klatt et al. \cite{klatt2022} found that PBGs may occur in finite-size structures without hyperuniformity. In the thermodynamic limit, however, they survive only in stealthy hyperuniform materials, a more restrictive form of hyperuniformity.
As for photonic crystals, 3D bicontinuous network geometries promote the formation of PBGs in disordered systems \cite{liew2011}. Complementary to the long-range description of hyperuniformity, Sellers et al. \cite{sellers2017} found that the similarity of the network's building blocks enhances PBGs. 
While their concept of local self-uniformity describes the similarity of network primitives (multiple bonds meeting in a vertex), Yang et al. \cite{yang2021} suggest that rings are the relevant local scatterers.

The dominant mechanism underlying structural color in disordered photonic networks with refractive indices too low to support PBGs remains unclear. Recently, we demonstrated that several 3D biophotonic networks are hyperuniform, suggesting that their structural color is a remnant of a PBG at a higher refractive index and thus a collective interference effect \cite{hemmann2026}. Alternatively to PBGs, structural color can arise from the excitation of local Mie and cavity resonances, enhanced by short-range interparticle interference. This picture explains the structural color of photonic glasses, correlated disordered assemblies of monodisperse spheres \cite{garcia2007, schertel2019}.
Photonic glasses readily generate brilliant blue colors, but saturated red colors are suppressed by cavity-like single-particle resonances and multiple scattering, which enhance high-frequency reflectance \cite{magkiriadou2014, shang2020, jacucci2020}.
Whereas the spheres in photonic glasses serve as well-defined local resonators, the corresponding resonant motifs in photonic networks have yet to be identified. Candidate resonators include the network's bonds, primitives, rings, and pores.

Here, we investigate the relevance of PBGs and local resonators on the structural coloration of 3D disordered networks. We apply an inverse method, originally developed in the context of electronic transport in disordered media~\cite{mukim2020}, to random photonic networks for the first time, aiming at retrieving their structural properties from reflectance spectra. We consider systems with a refractive index of $n=1.5$, typical for biophotonic materials \cite{vogler-neuling2023} and use a dataset of computer-generated disordered networks, whose short-range order is controlled by a single disorder parameter $\beta$ (to be subsequently defined).
We demonstrate that high disorder in the network's rings broadens its reflectance peak. By comparing the reflectance spectra of different network types, characterized by their coordination statistics, we identify rings and pores as the relevant local scatterers whose sizes shift the reflectance spectrum. We show that our inverse method can detect spectral shifts as well as broadening and, therefore, infer the mean and standard deviation of ring radii solely from a network's reflectance.

Furthermore, we apply our inverse method to the biophotonic network that gives rise to blue structural color in the \textit{Pachyrhynchus congestus mirabilis} weevil \cite{djeghdi2022} and find that diamond- and \textbf{ctn}-like networks best reproduce the biological spectrum by matching pore and ring sizes. In contrast, we find no significant effect of hyperuniformity and primitive similarity on the reflectance spectra. This suggests that blue structural color is not a photonic band gap effect. The importance of network pores rather suggests an analogy to an inverse photonic glass, explaining the predominant blue-to-green colors in computer-generated and biological networks.

\section*{Results and Discussion}

\begin{figure*}[htbp]
    \centering
    \begin{tikzpicture}

        \def\imgWLeft{0.2\textwidth}  
        \def\imgWRight{0.36\textwidth} 
        
        \def\colSep{0.1cm}        
        \def\rowSep{0.4cm}       
        \def\blockSep{1.0cm}      
        
        \tikzset{
            image/.style={inner sep=0pt, outer sep=0pt, anchor=north west},
            insetLabel/.style={
                anchor=north west,
                fill=white,
                fill opacity=0.8,
                text opacity=1,
                inner sep=3pt,
                font=\sffamily\bfseries 
            },
            arrowLabel/.style={
                font=\sffamily\footnotesize 
            }
        }

        \node[image] (A) at (0,0) {\includegraphics[width=\imgWLeft]{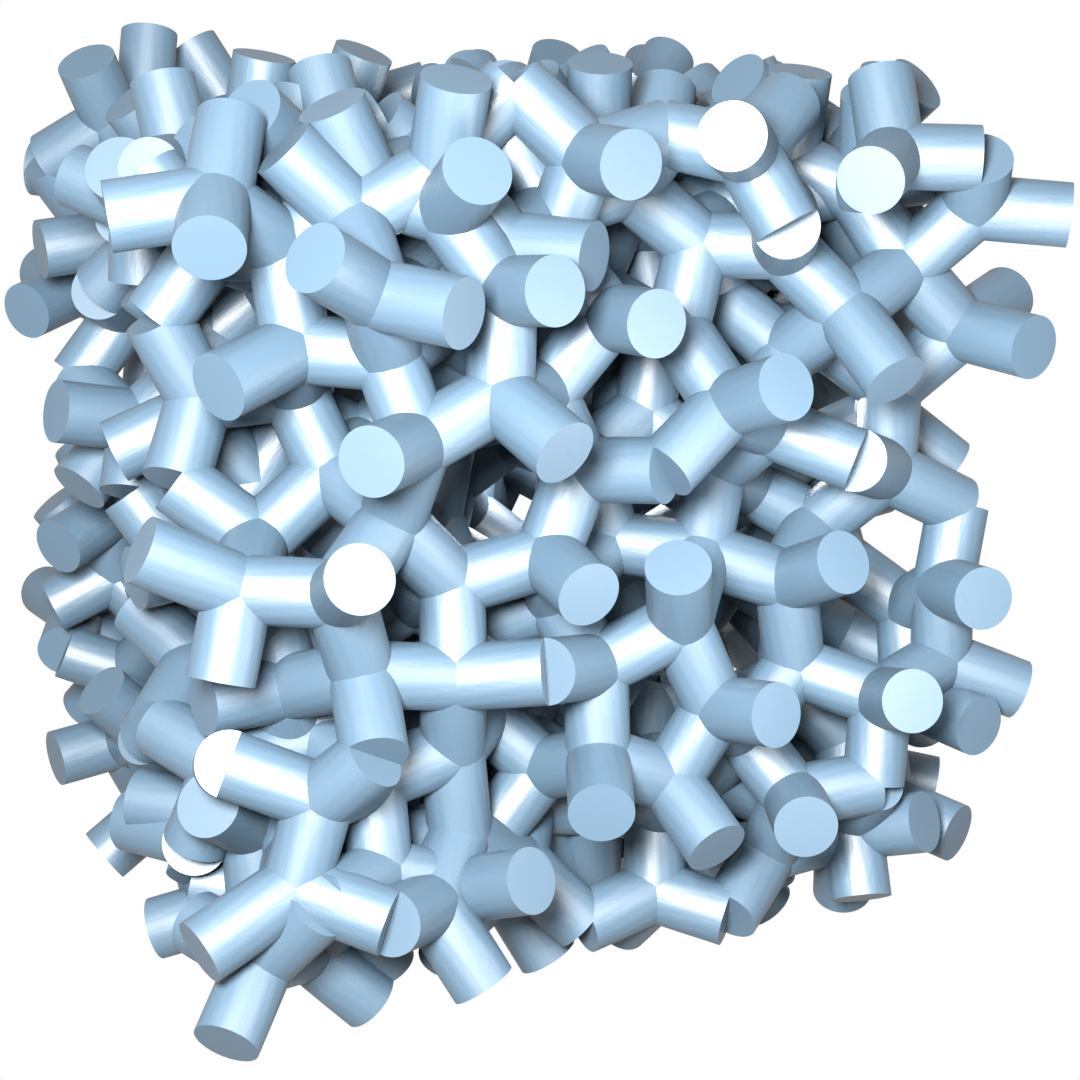}};
        \node[insetLabel] at (A.north west) {A};

        \node[image] (C) at ([xshift=\colSep]A.north east) {\includegraphics[width=\imgWLeft]{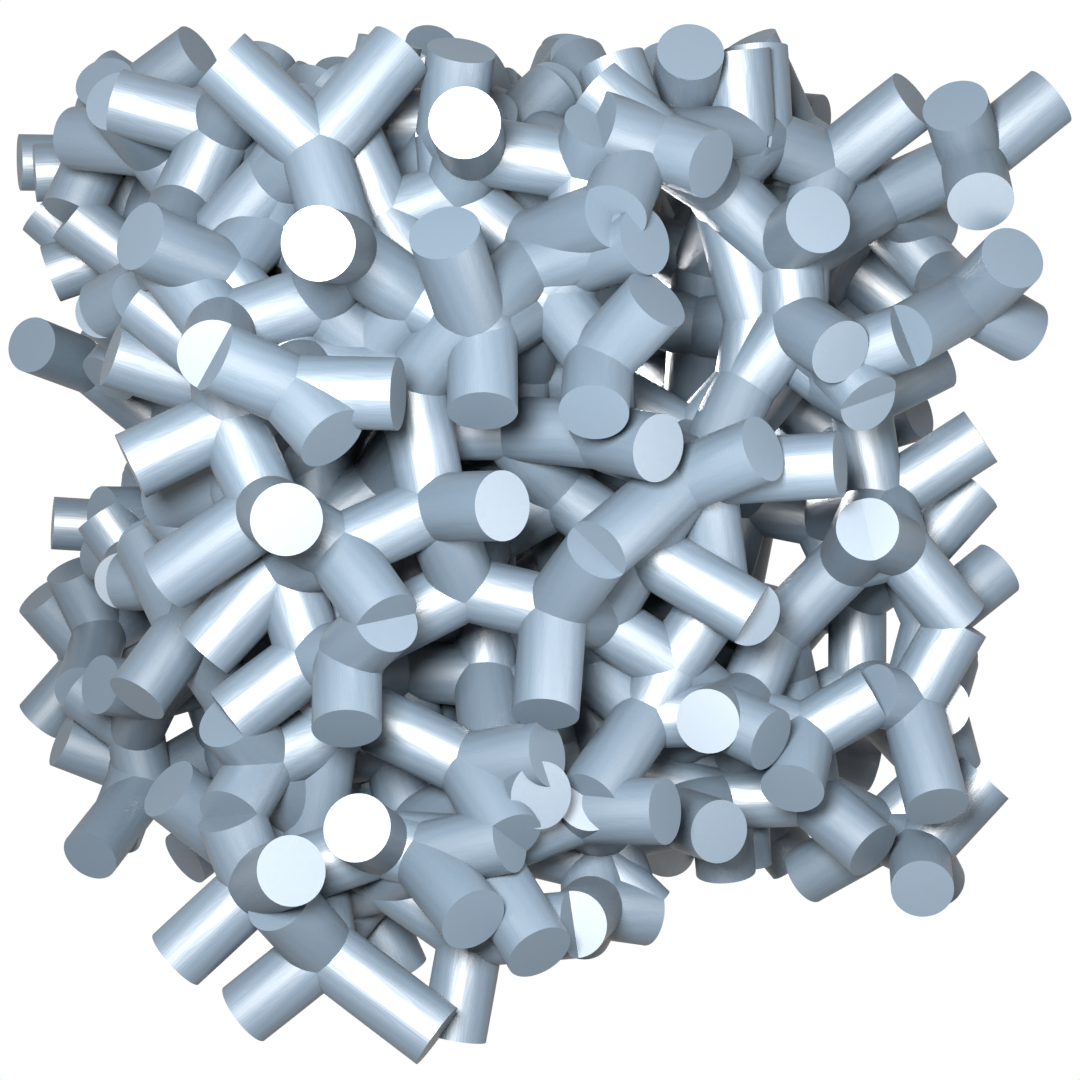}};
        \node[insetLabel] at (C.north west) {C};

        \node[image] (E) at ([xshift=\blockSep]C.north east) {\includegraphics[width=\imgWRight]{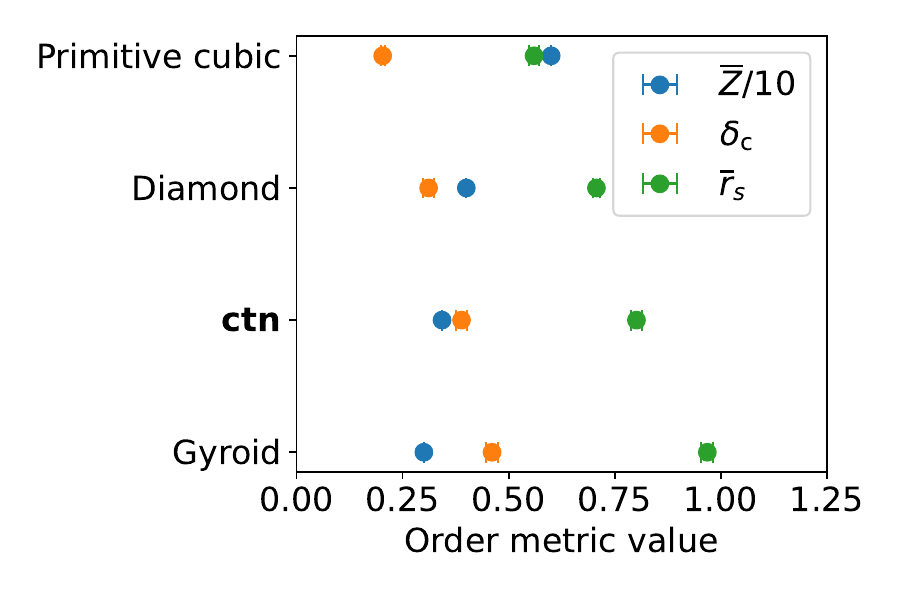}};
        \node[insetLabel] at ([xshift=-0.3cm]E.north west) {E};

        \node[image] (B) at ([yshift=-\rowSep]A.south west) {\includegraphics[width=\imgWLeft]{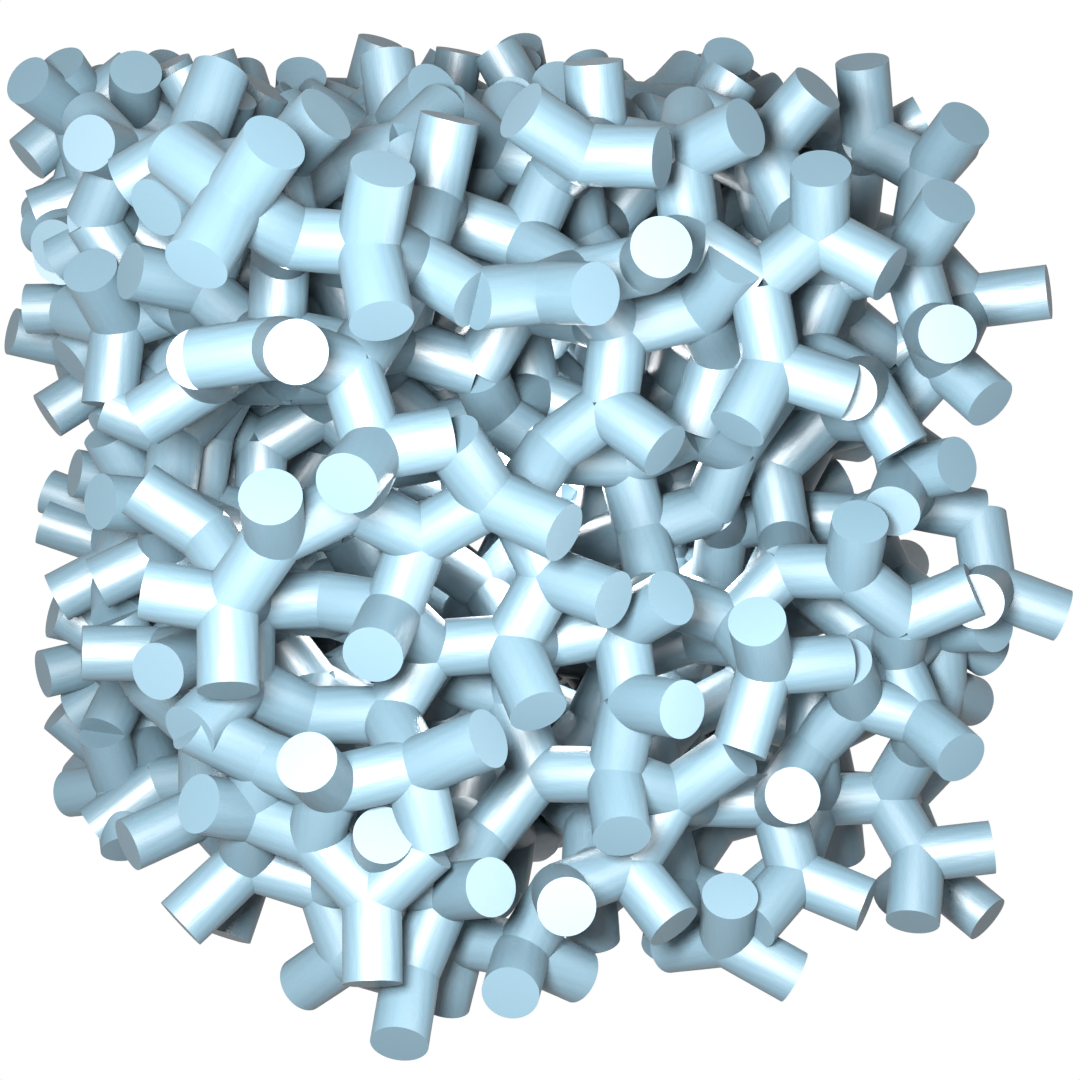}};
        \node[insetLabel] at (B.north west) {B};

        \node[image] (D) at ([xshift=\colSep]B.north east) {\includegraphics[width=\imgWLeft]{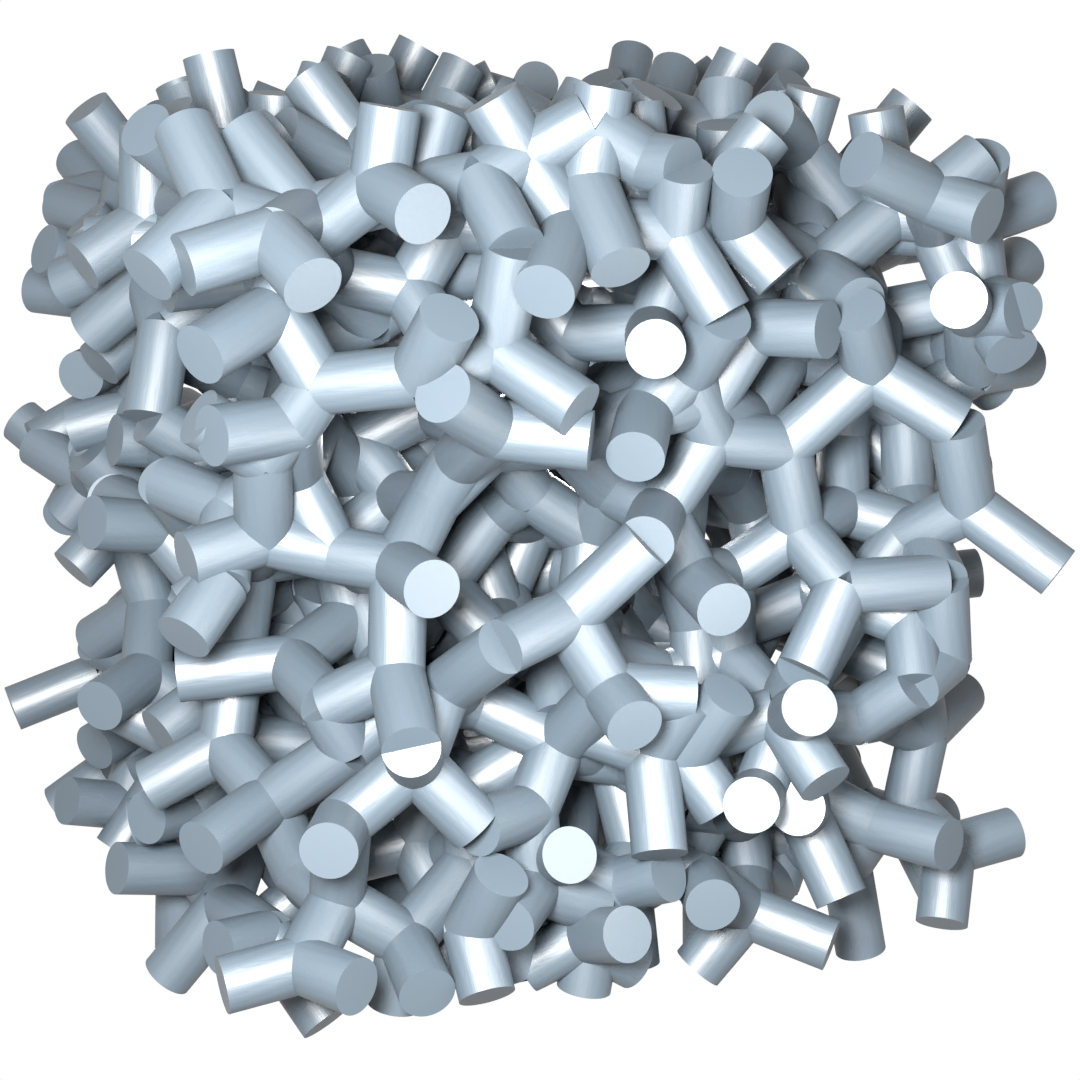}};
        \node[insetLabel] at (D.north west) {D};

        \node[image] (F) at ([xshift=\blockSep]D.north east) {\includegraphics[width=\imgWRight]{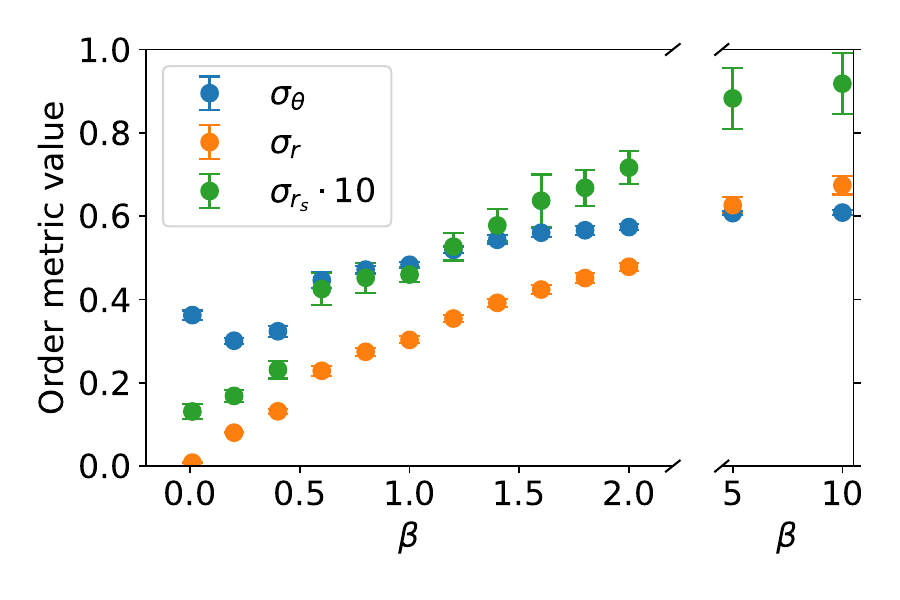}};
        \node[insetLabel] at ([xshift=-0.3cm]F.north west) {F};

        \draw[->, thick, >=stealth] 
            ([yshift=-0.3cm]B.south west) -- ([yshift=-0.3cm]D.south east) 
            node[midway, below, arrowLabel] {$\beta$, short-range disorder};

        \draw[->, thick, >=stealth] 
            ([xshift=-0.3cm]B.south west) -- ([xshift=-0.3cm]A.north west) 
            node[midway, above, rotate=90, arrowLabel] {$\overline{Z}$, network type};

    \end{tikzpicture}
    \caption{The effect of network type and the disorder parameter $\beta$ on short-range order metrics and color appearance. (A,B) Diamond- and ctn-like networks generated at beta = 0.2. (C,D) Corresponding networks generated at beta = 0.8. Diamond-like networks have $\overline{Z}=4$, ctn-like networks $\overline{Z}=3.4$. The rendered colors in (A)-(D) correspond to the simulated optical response for an equilibrium bond length $d=\SI{160}{\nano\meter}$. (E) An increase in mean coordination scales down the critical pore size $\delta_\mathrm{c}$ and the mean ring radius $\overline{r}_s$ for networks generated at $\beta=0.2$. (F) As the network generation parameter $\beta$ of diamond-like networks increases, the bond length standard deviation $\sigma_r$ and the ring radius standard deviation $\sigma_{r_s}$ grow. The bond angle standard deviation $\sigma_\theta$ follows an opposing trend for small $\beta$, as expected from Equation~\ref{eqn:keating_generalized}.}
    \label{fig:networks_order_metrics}
\end{figure*}
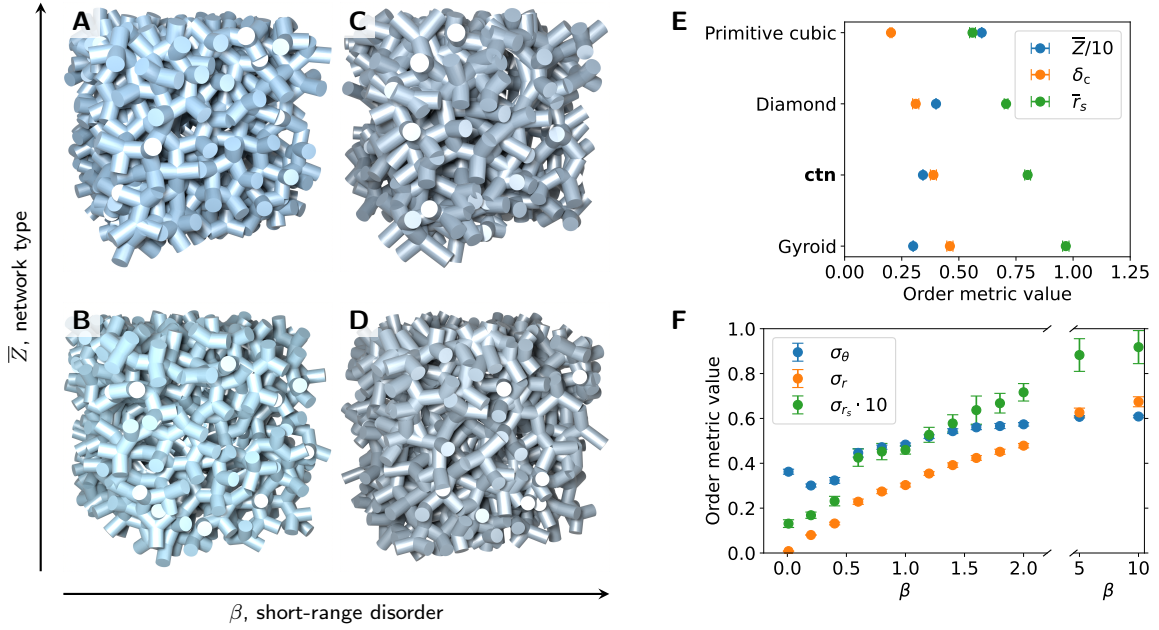

\subsection*{Network Type and $\bm{\beta}$ Affect Order Metrics}
We use an extended Wooten-Weaire-Winer algorithm \cite{wooten1985, hemmann2026} to generate a dataset of 3D disordered networks. Disorder is introduced into an initially crystalline network by proposing a bond-switch move which is accepted with the Metropolis probability $P_\mathrm{accept}=\min \left[1, \exp \left( (E_\mathrm{f}-E_\mathrm{i})/T \right) \right]$. $T$ is a temperature parameter, $E_\mathrm{i}$ is the initial, and $E_\mathrm{f}$ the final strain energy \cite{metropolis1953}. We use a modified Keating strain energy \cite{hemmann2026}
\begin{equation}
    E= E_r + \beta E_\theta
    \quad ,
    \label{eqn:keating_generalized}
\end{equation}
where $E_r$ is a bond-stretching and $E_\theta$ a bond-bending energy. $E_r$ is approximately harmonic around the equilibrium bond length $d$, which is the intrinsic length scale, while $E_\theta$ promotes bond repulsion. The bond-bending constant $\beta$ weights the second term and, thus, controls whether disorder is rather introduced into bond angles (small $\beta$) or lengths (large $\beta$). 

The networks' structural characteristics are governed by the choice of the initial network and the disorder parameter $\beta$ (Figure~\ref{fig:networks_order_metrics}). Since the bond-switch leaves the coordination number invariant, the coordination mean $\overline{Z}$ and standard deviation $\sigma_Z$ are set by the initial network. Figures \ref{fig:networks_order_metrics}A-B illustrate disordered networks generated at $\beta=0.2$, starting from a cubic diamond network with $\overline{Z}=4$, and from the \textbf{ctn} network with $\overline{Z}=3.4$ \cite{rcsr_ctn2025}. The mean coordination numbers anticorrelate with the critical pore radius $\delta_\mathrm{c}$ and the mean ring radius $\overline{r}_s$ (Figure~\ref{fig:networks_order_metrics}E) \cite{hemmann2026}. This trend is supported by the inclusion of primitive-cubic-like networks with $\overline{Z}=6$ and gyroid-like networks with $\overline{Z}=3$.
All lengths, including $\delta_\mathrm{c}$ and $\overline{r}_s$, are given in units of the equilibrium bond length $d$. 

Figures \ref{fig:networks_order_metrics}C-D show a diamond- and a \textbf{ctn}-like network, generated at a higher bond-bending constant $\beta=0.8$. As expected from Equation~\ref{eqn:keating_generalized}, for $\beta\lesssim 0.2$, an increase in $\beta$ stiffens the angles and thus reduces the bond angle standard deviation $\sigma_\theta$ while raising the bond length standard deviation $\sigma_r$ and the related ring radius standard deviation $\sigma_{r_s}$ (Figure~\ref{fig:networks_order_metrics}F). For $\beta\gtrsim 0.2$, the angle stiffening prevents the network generation algorithm from reaching configurations of high angle order, and $\sigma_\theta$ increases with $\beta$. While the mentioned short-range order metrics correlate with the network type and $\beta$, long-range order metrics, such as the hyperuniformity metric $\alpha$, are not directly controlled by $\overline{Z}$ or $\beta$.
Hyperuniform systems exhibit $\alpha>0$ and are uniform at large or infinite length scales \cite{torquato2018}. In our network dataset, only diamond-like networks with $\beta=0.01$ are, on average, hyperuniform with $\alpha=0.23 \pm 0.30$.

\subsection*{Inverse Determination of Network Type and Short-Range Order}


We now investigate the effect of the discussed order metrics on the networks' reflectance spectra using finite-difference time-domain simulations and an inverse method. 
To assign a volume to the mathematical networks, we decorate the bonds with cylinders with refractive index $n_\mathrm{m}=1.5$ and radius $R=0.39$, where diamond- and \textbf{ctn}-like networks exhibit volume fractions of $\phi\approx 0.4$. According to Equation~\ref{eqn:volume_fraction_pbg}, this volume fraction is expected to produce the strongest color response for the given refractive index. 

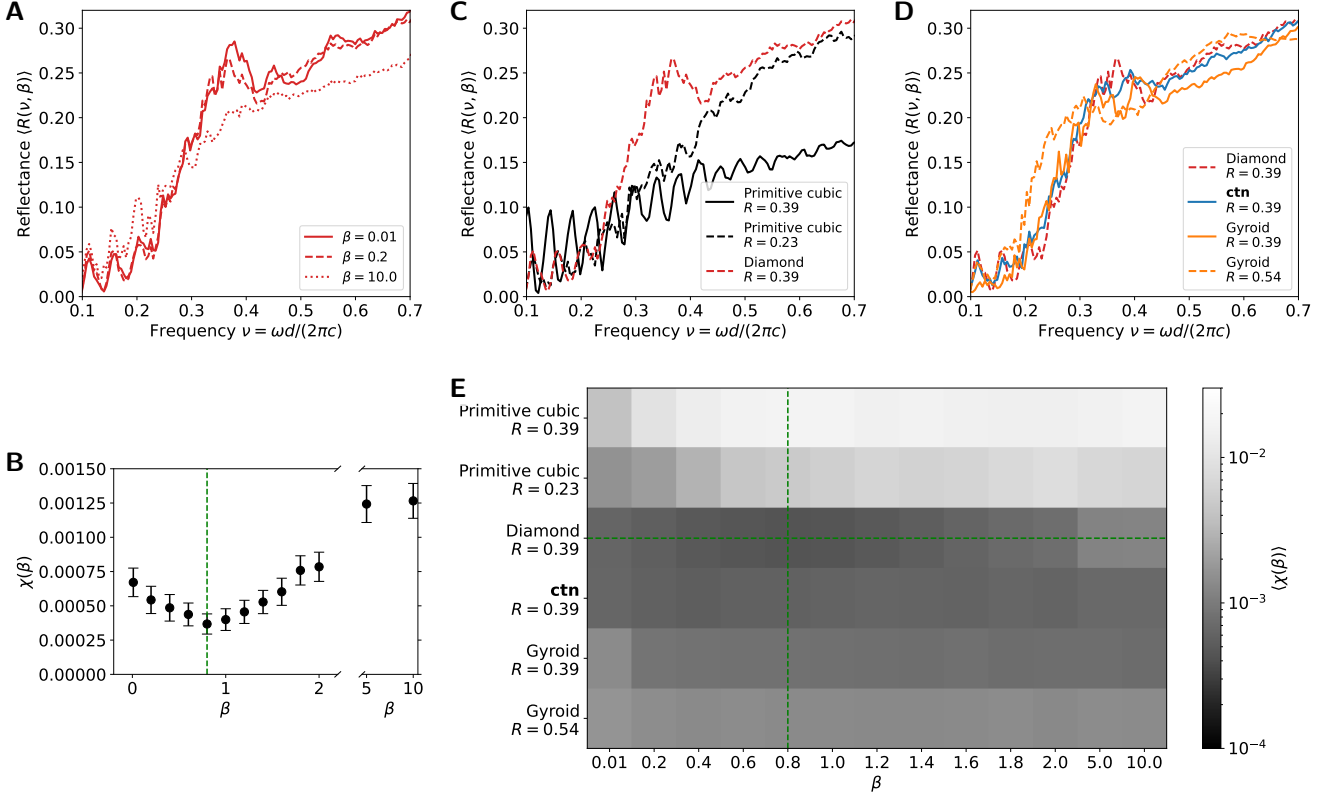
\begin{figure*}[htbp]
    \centering
    \begin{tikzpicture}

        \def\imgWSmall{0.33\textwidth} 
        \def\colSep{0.0\textwidth}    
        \def\rowSep{0.1cm}             
        
        \def\imgWLarge{0.66\textwidth} 

        \tikzset{
            image/.style={inner sep=0pt, outer sep=0pt},
            insetLabel/.style={
                anchor=north west,
                fill=white,
                fill opacity=0.8,
                text opacity=1,
                inner sep=3pt,
                font=\sffamily\bfseries
            }
        }

        \node[image, anchor=north west] (A) at (0,0) {\includegraphics[width=\imgWSmall]{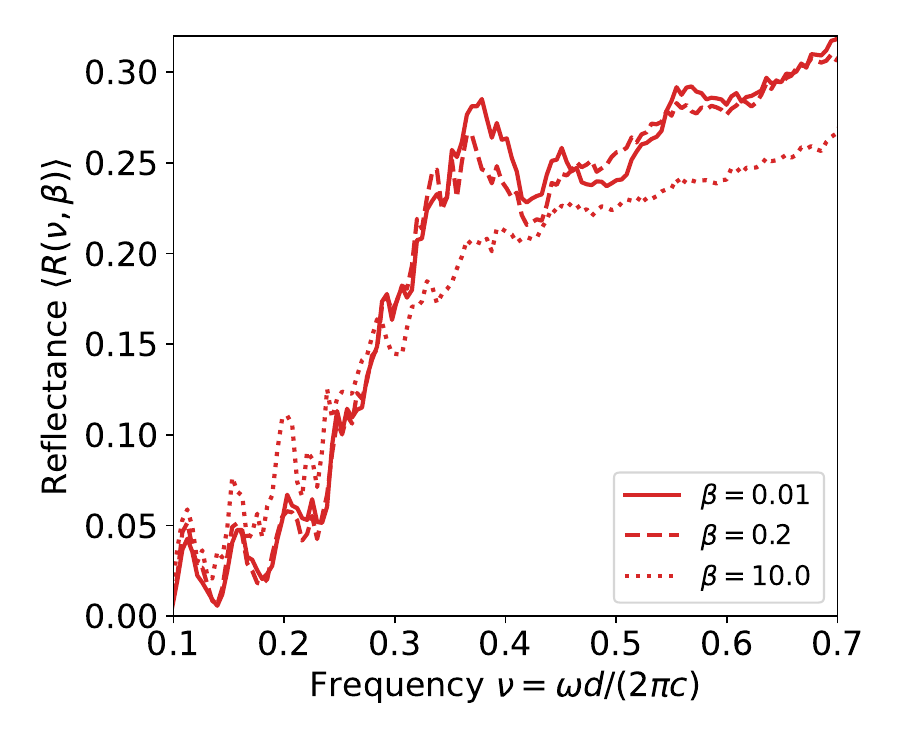}};
        \node[insetLabel] at (A.north west) {A};

        \node[image, anchor=north west] (C) at ([xshift=\colSep]A.north east) {\includegraphics[width=\imgWSmall]{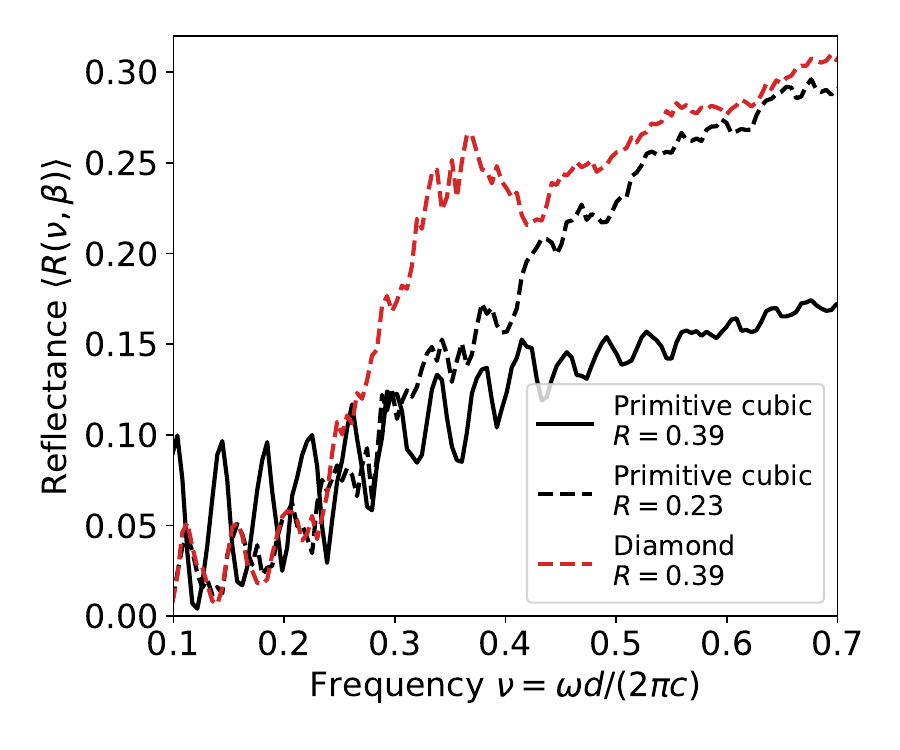}};
        \node[insetLabel] at (C.north west) {C};

        \node[image, anchor=north west] (D) at ([xshift=\colSep]C.north east) {\includegraphics[width=\imgWSmall]{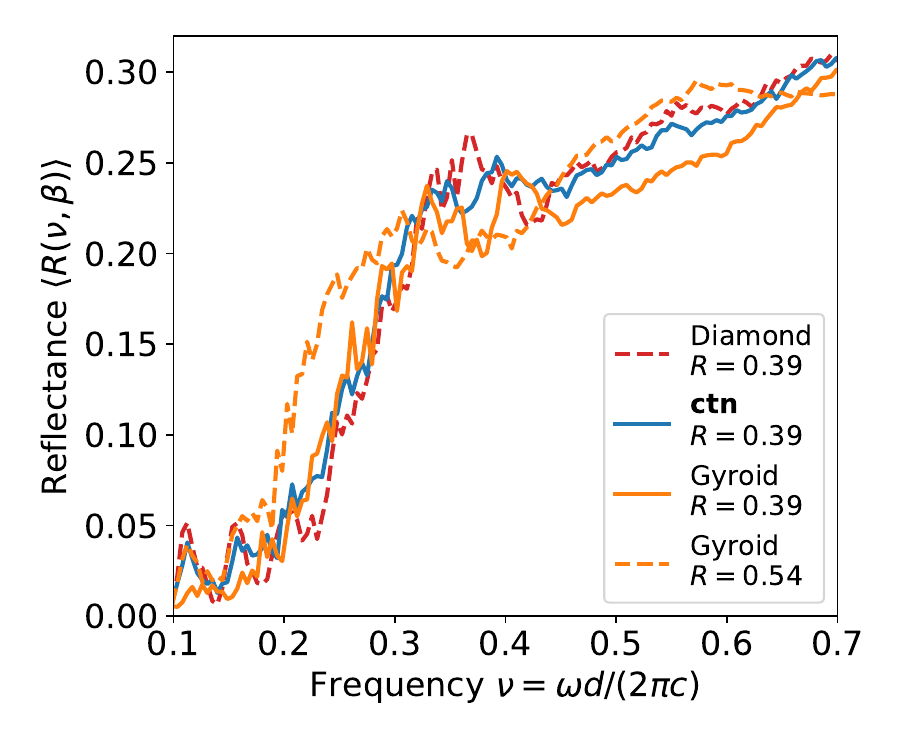}};
        \node[insetLabel] at (D.north west) {D};

        \node[image, anchor=north west] (E) at ([yshift=-\rowSep]C.south west) {\includegraphics[width=\imgWLarge]{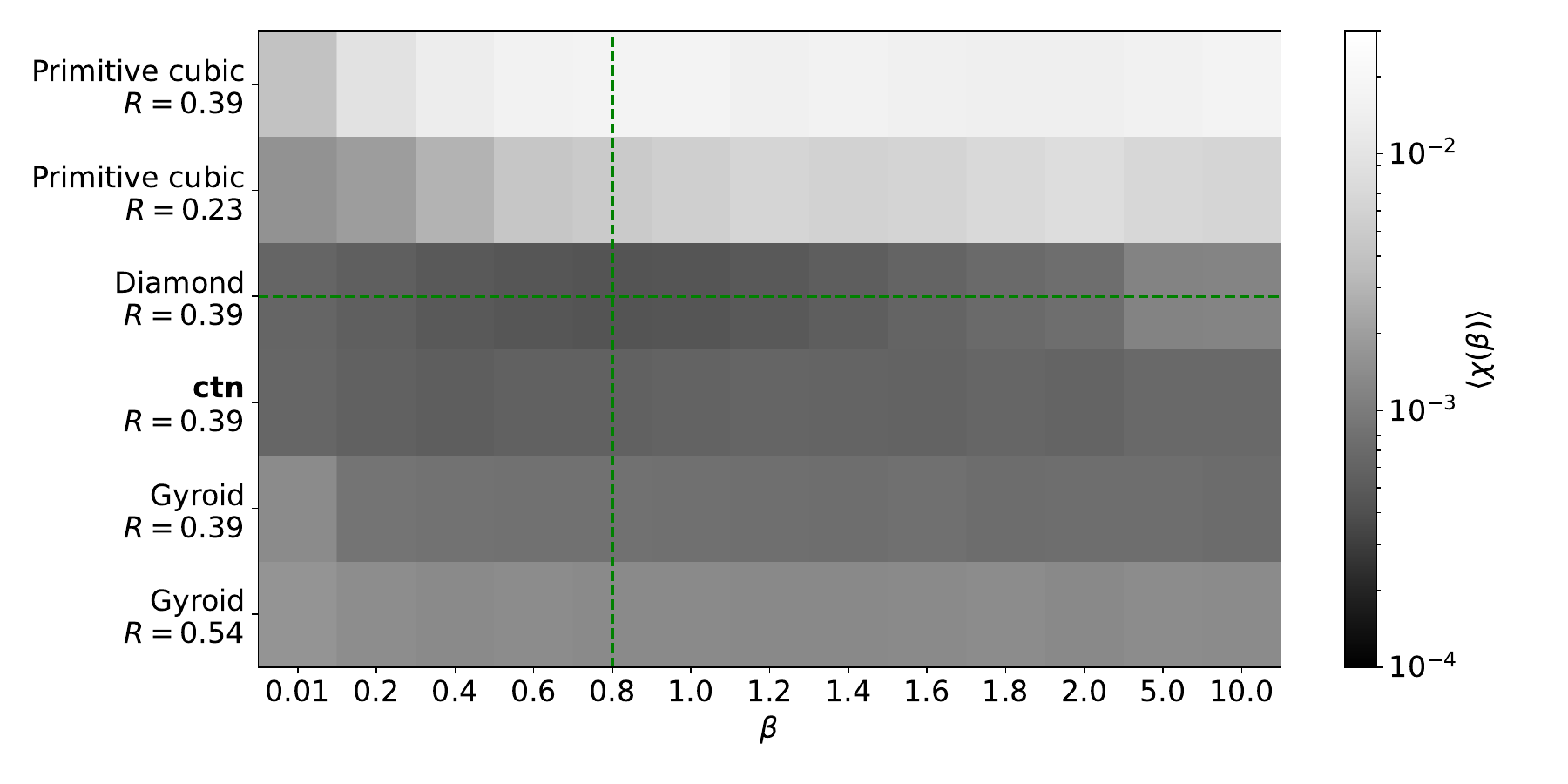}};
        \node[insetLabel] at (E.north west) {E};

        \node[image, anchor=east] (B) at ([xshift=-\colSep]E.west) {\includegraphics[width=\imgWSmall]{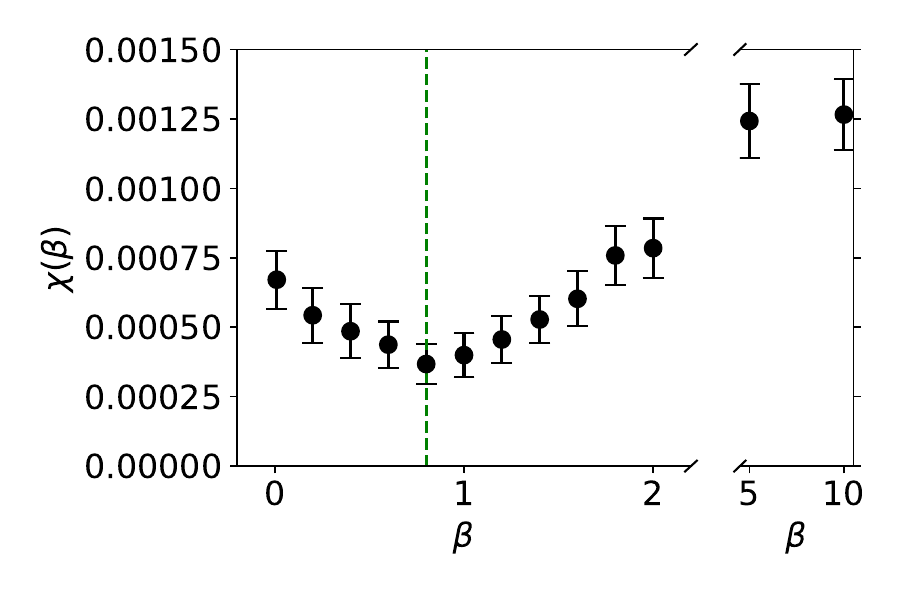}};
        \node[insetLabel] at (B.north west) {B};

    \end{tikzpicture}
    \caption{The disorder parameter $\beta$ primarily affects the spectral shape of the reflectance, while the network type induces a frequency shift. These effects are detected by the misfit function, which can be used to determine $\beta$ and the network type of an unknown spectrum. (A) Average reflectance spectra for diamond-like networks with $R=0.39$ and three values of $\beta$. (B) Misfit function comparing a diamond-like test network with $\beta=0.8$ to diamond-like networks of all $\beta$ values. Average reflectance spectra of different network types with (C) $\overline{Z}\geq 4$ and (D) $\overline{Z} \leq 4$. All networks in (C) and (D) were generated at $\beta=0.2$. For primitive-cubic- and gyroid-like networks, we vary the bond radius to bring the volume fraction to $\phi\approx 0.4$, close to the $\phi$ values of diamond- and \textbf{ctn}-like networks at $R=0.39$. (E) Average misfit functions comparing diamond-like networks with $\beta=0.8$ to all other network types and $\beta$ values.}
    \label{fig:reflectance_misfit_fct}
\end{figure*}

Figure~\ref{fig:reflectance_misfit_fct}A displays the average reflectance spectra of diamond-like networks with different $\beta$ values.
At low frequencies $\nu$, the reflectance is weak because light averages over the network structure with an effective refractive index only slightly above unity. The oscillations at $\nu \lesssim 0.2$ arise from Fabry-Pérot resonances of the finite-thickness slab.
As the frequency enters the resonant regime at $\nu \gtrsim 0.2$, the reflectance increases and peaks near $\nu \approx 0.3$. The peak broadens with increasing $\beta$, consistent with the enhanced bond-length and ring disorder shown in Figure~\ref{fig:networks_order_metrics}F. Interpreting these structural motifs as local resonators, size disorder broadens the distribution of resonance frequencies. This explanation is in line with the argument that primitive self-uniformity enhances photonic band gaps and, thus, creates a more peaked reflectance spectrum \cite{sellers2017}. Since diamond-like networks at $\beta=0.01$ are, on average, hyperuniform, their pronounced reflectance peak may be a remnant of a PBG at higher dielectric contrast.

We use the relation between short-range disorder and the reflectance peak broadening to inversely determine the disorder parameter $\beta$ of an unknown network. Mukim et al. \cite{mukim2020}
introduced a versatile, quite general method to infer the structural disorder strength $\beta$ of an unknown disordered system from transport properties, such as conductance or reflectance. Based on the ergodicity hypothesis, they argue that the actual, correct value of $\beta$ is found by minimizing the so-called misfit function
\begin{equation*}
\chi \left(\beta\right)=\int_{\nu_\mathrm{min}}^{\nu_\mathrm{max}}
d\nu \,
\left[ R(\nu) - \langle R(\nu, \beta) \rangle \right]^2,
\end{equation*}
which compares the target reflectance spectrum $R(\nu)$ to the ensemble averaged spectrum $\langle R(\nu, \beta) \rangle$.
The hypothesis avails of the fact that averaging the signal over a continuously varying parameter $\nu$ yields the same result as averaging over an ensemble of many different disorder configurations.
This method has successfully been applied to infer structural disorder \cite{mukim2020, mukim2022} and the electron-photon coupling strength \cite{macedo2026} from the transmission spectra of low-dimensional quantum systems.
Recently, this inverse tool has also been shown to infer stealthy hyperuniform correlations in model Hamiltonians~\cite{vanoni2026effective} from quantum transport~\cite{costa2026inferring}.
Here we introduce this inverse methodology to photonics to demonstrate that the minimum of the misfit function correctly predicts the unknown disorder parameter $\beta$ from the reflectance of diamond-like networks (Figure~\ref{fig:reflectance_misfit_fct}B). Thus, using the relation between $\beta$ and the short-range order metrics plotted in Figure~\ref{fig:networks_order_metrics}F, the inverse method discloses relevant information on structural disorder, including ring-radius disorder, using only the reflectance signal.

We now investigate how the changes in coordination numbers, ring and pore sizes between different network types affect their reflectance spectra. For bond radii $R=0.39$, the spectrum of primitive-cubic-like networks differs significantly from those of the other network types (Figures~\ref{fig:reflectance_misfit_fct}C-D). It is much flatter and exhibits dominant Fabry-Pérot resonances at small frequencies. Due to the higher bond density of primitive-cubic-like networks, at $R=0.39$, they exhibit volume fractions of $\phi\approx0.72$, much higher than $\phi\approx0.4$, as present for diamond- and \textbf{ctn}-like networks. At this high volume fraction, the network is not bicontinuous, but contains closed air pores. Bicontinuity is known to favor PBGs at high refractive index contrast $n_\mathrm{m}/n_\mathrm{b}$, and consequently stronger reflectance peaks at lower $n_\mathrm{m}/n_\mathrm{b}$ \cite{economou1993, liew2011}. To take into account the volume fraction and restore bicontinuity, we reduce the bond radius in primitive-cubic-like networks to $R=0.23$, yielding $\phi\approx0.47$. The resulting spectrum shows a clearer increase in reflection with increasing frequency, indicating that the local resonators have been restored (Figure~\ref{fig:reflectance_misfit_fct}C). The alignment of the weak Fabry-Pérot fringes at $\nu \leq 0.2$ suggests similar effective refractive indices among volume-fraction-matched systems.
Comparing the reflectance peak of primitive-cubic-like networks with $R=0.23$ to diamond-, \textbf{ctn}-, and gyroid-like networks at $R=0.39$, we notice a redshift with decreasing mean coordination number (Figure~\ref{fig:reflectance_misfit_fct}C-D).
We attribute this redshift to the increasing ring and pore sizes (Figure~\ref{fig:networks_order_metrics}F). By interpreting rings and pores as local resonators, the resonance frequency (in units of $c/d$) increases as the ring and pore size (in units of $d$) decrease.

The redshift is pronounced between primitive-cubic-like networks and the other network types, but comparatively small among diamond-, \textbf{ctn}-, and gyroid-like networks. We attribute this behavior to a second trend that partially compensates for the influence of ring and pore size: the bond density, and thus the volume fraction, decreases with increasing $\overline{Z}$. For $R=0.39$, we obtain $\phi\approx0.45$ for diamond-like, $\phi\approx0.36$ for \textbf{ctn}-~, and $\phi\approx0.23$ for gyroid-like networks. Although an effective refractive index cannot be rigorously defined for resonant photonic networks, a lower volume fraction is nevertheless expected to reduce the average refractive index. The resulting decrease in optical path length shifts the resonances to higher frequencies, counteracting the trend associated with ring and pore size. For gyroid-like networks, we compensate for the reduced volume fraction by increasing the bond radius to $R=0.54$, yielding $\phi\approx0.39$. With the volume-fraction offset removed, the red shift associated with the large rings and pores becomes pronounced (Figure~\ref{fig:reflectance_misfit_fct}D). Consistent with the primitive-cubic-like networks, volume-fraction matching also aligns the Fabry-Pérot resonances in the homogenized regime.

Figure~\ref{fig:reflectance_misfit_fct}E illustrates that the inverse method can indeed determine both the network type and the short-range order metrics of an unknown network. We plot the average misfit function $\langle\chi(\beta)\rangle$, obtained by averaging the misfit functions of 10 test networks. Besides the spectral broadening related to $\beta$, the average misfit function detects the reflectance shift caused by differences in rings and pore sizes.

\subsection*{Application to a Blue Biophotonic Network}
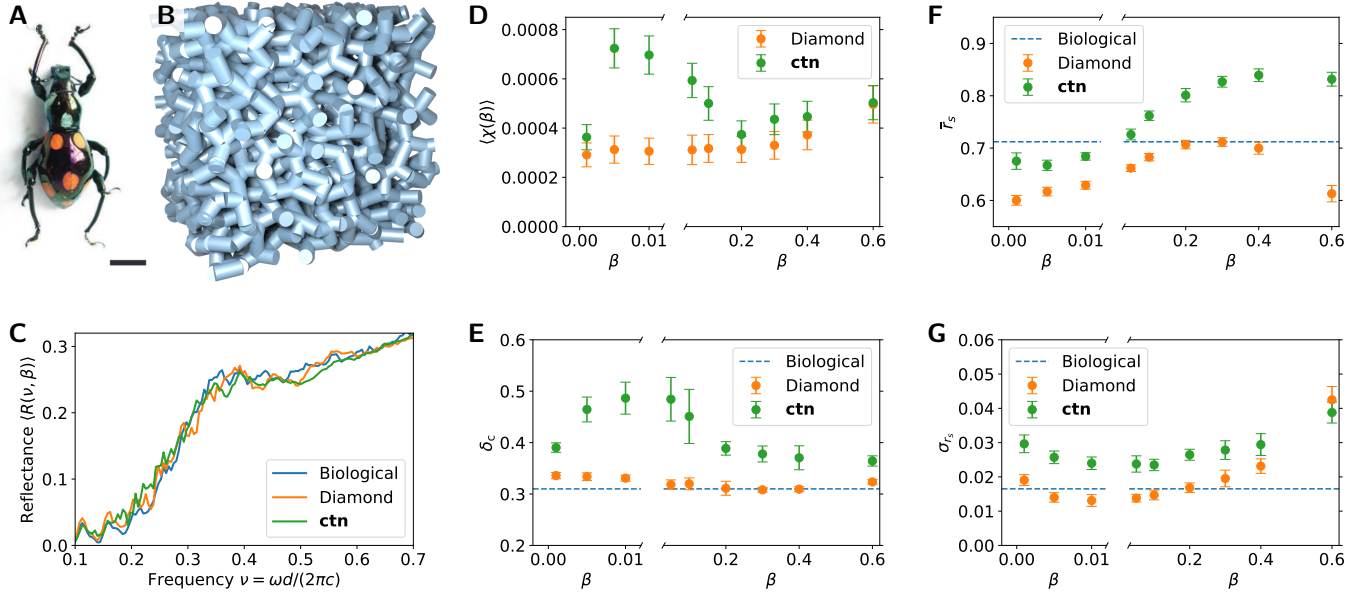
\begin{figure*}[htbp]
    \centering
    \begin{tikzpicture}

        \def\imgWA{0.11\textwidth}     
        \def\imgWB{0.22\textwidth}     
        \def\imgWMain{0.33\textwidth}  
        
        \def\colSepSub{0.0cm}          
        \def\colSep{0.2cm}             
        \def\rowSep{0.3cm}             

        \tikzset{
            image/.style={inner sep=0pt, outer sep=0pt},
            insetLabel/.style={
                anchor=north west,
                fill=white,
                fill opacity=0.8,
                text opacity=1,
                inner sep=3pt,
                font=\sffamily\bfseries
            }
        }

        \node[image, anchor=west] (A) at (0,0) {\includegraphics[width=\imgWA]{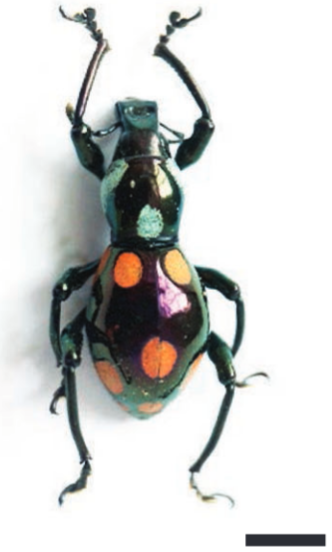}};

        \node[image, anchor=west] (B) at ([xshift=\colSepSub]A.east) {\includegraphics[width=\imgWB]{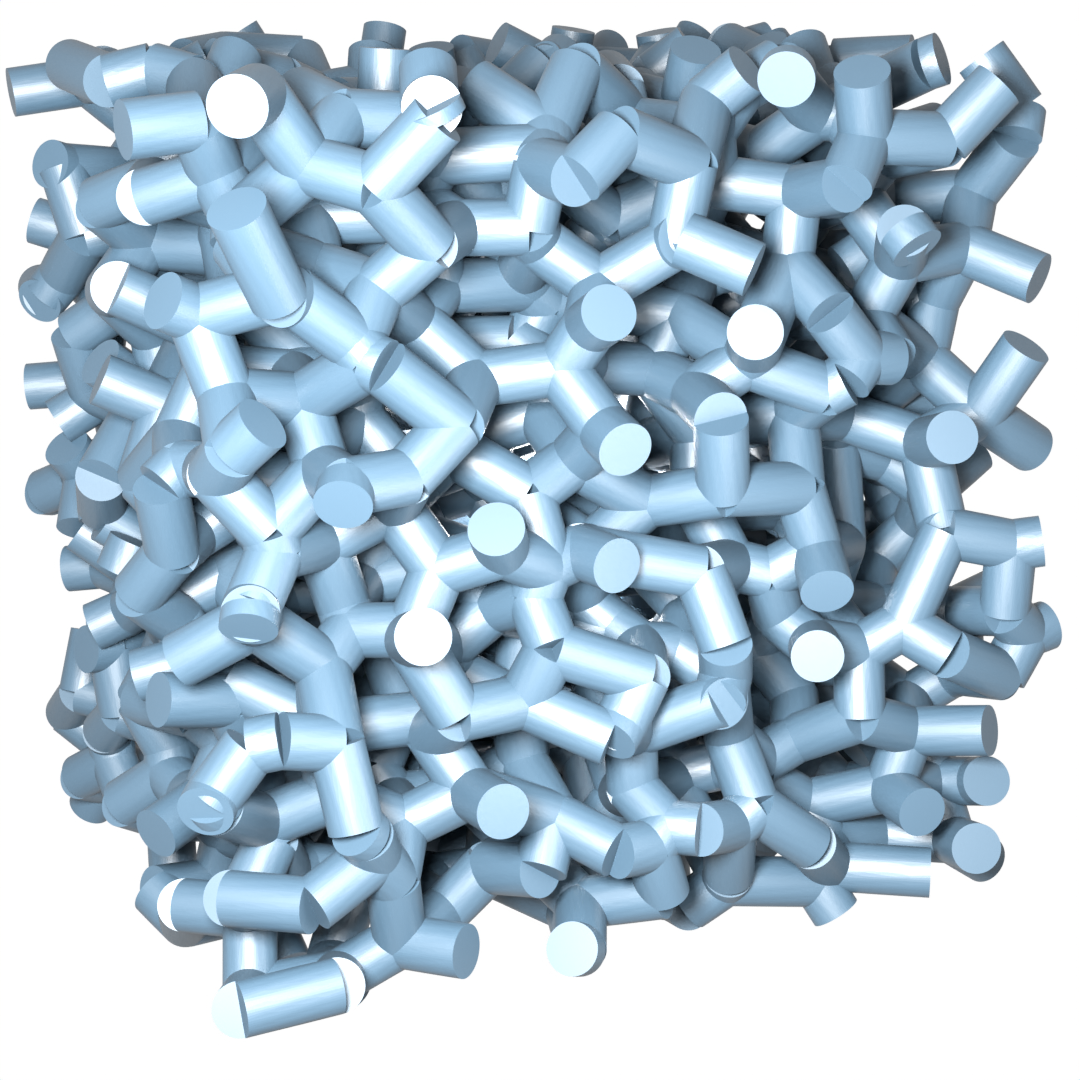}};

        \node[image, anchor=west] (D) at ([xshift=\colSep]B.east) {\includegraphics[width=\imgWMain]{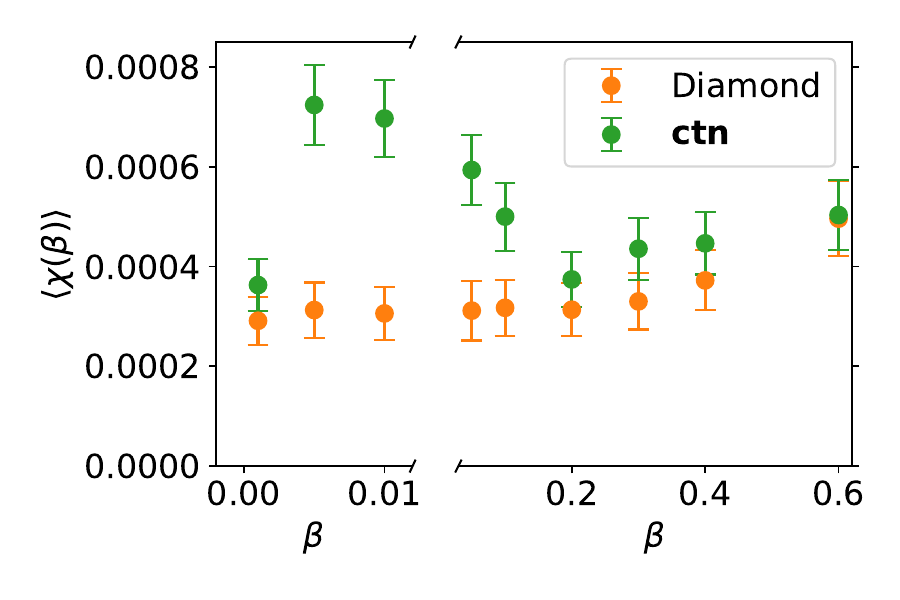}};

        \node[image, anchor=west] (F) at ([xshift=\colSep]D.east) {\includegraphics[width=\imgWMain]{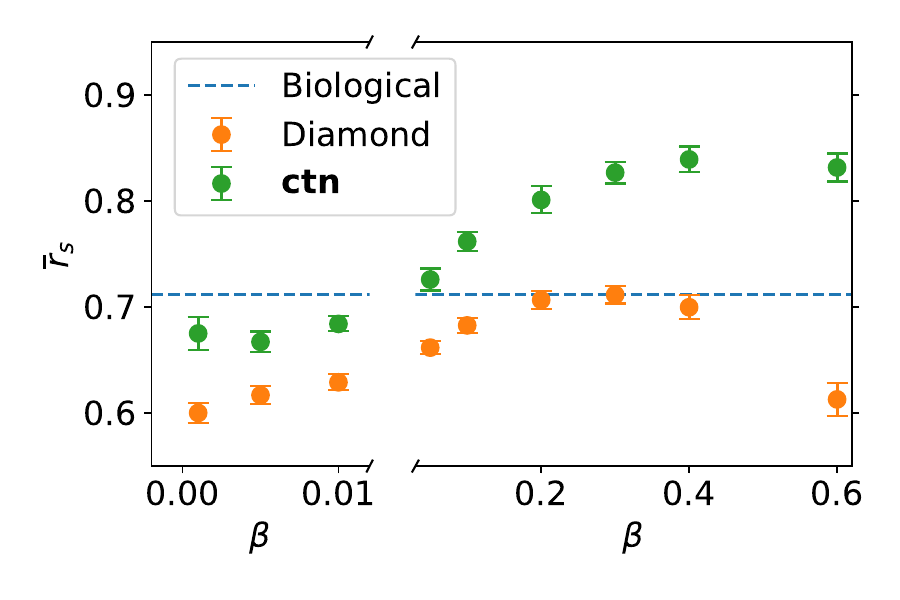}};

        \coordinate (TopEdgeRow1) at (current bounding box.north);

        \node[insetLabel] at (A.west |- TopEdgeRow1) {A};
        \node[insetLabel] at (B.west |- TopEdgeRow1) {B};
        \node[insetLabel] at (D.west |- TopEdgeRow1) {D};
        \node[insetLabel] at (F.west |- TopEdgeRow1) {F};

        \coordinate (BottomRow1) at (current bounding box.south);

        \node[image, anchor=north west] (C) at ([yshift=-\rowSep] A.west |- BottomRow1) {\includegraphics[width=\imgWMain]{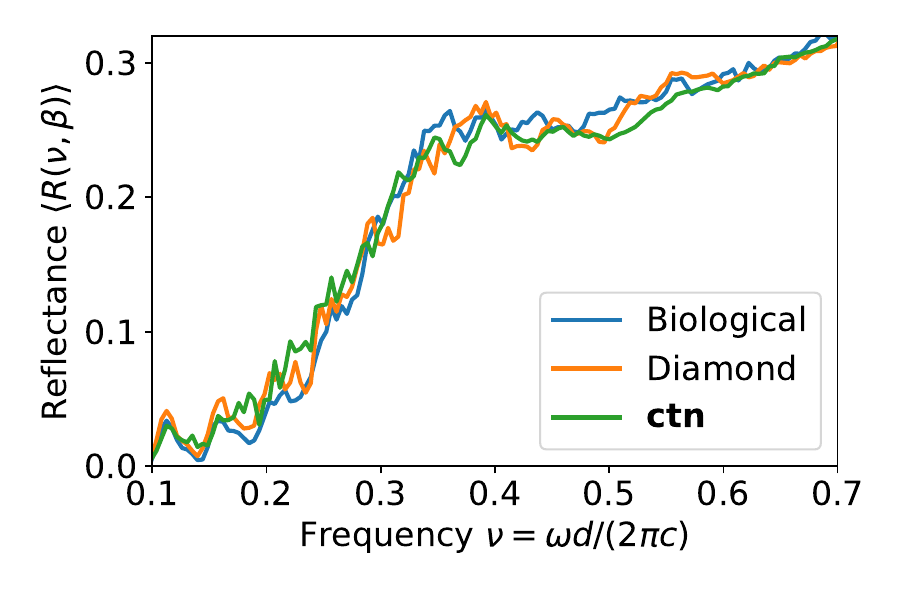}};

        \node[image, anchor=north west] (E) at (D.west |- C.north) {\includegraphics[width=\imgWMain]{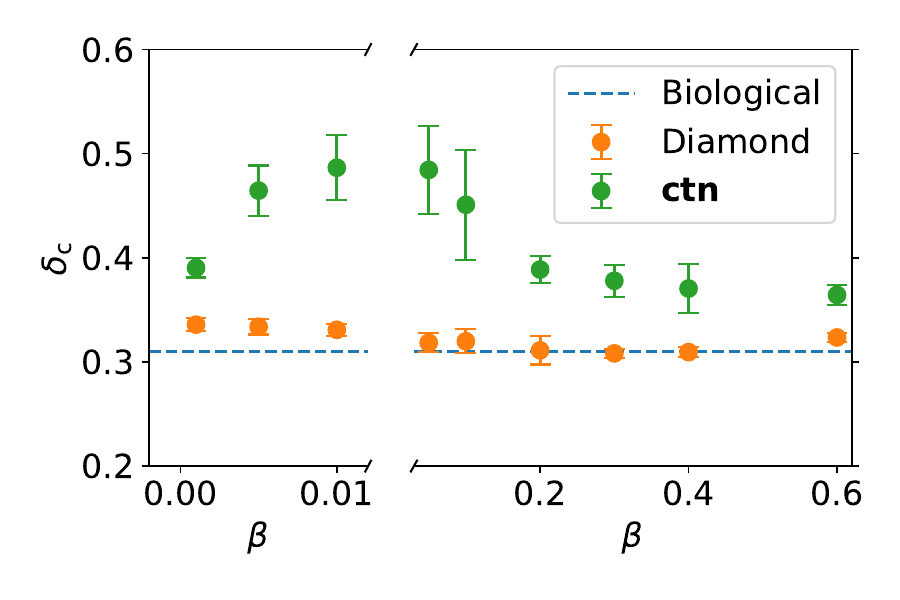}};

        \node[image, anchor=north west] (G) at (F.west |- C.north) {\includegraphics[width=\imgWMain]{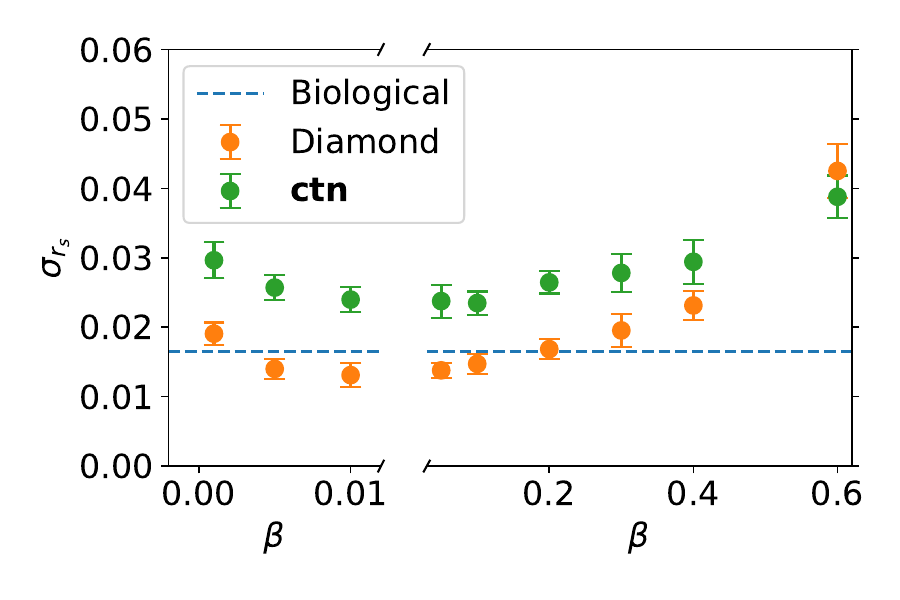}};

        \node[insetLabel] at (C.north west) {C};
        \node[insetLabel] at (E.north west) {E};
        \node[insetLabel] at (G.north west) {G};

    \end{tikzpicture}
    \caption{The reflectance spectrum of the biophotonic network exhibiting blue structural color resembles the spectra of diamond- and \textbf{ctn}-like networks when their pore and ring radii match. (A) Photograph of a \textit{Pachyrhynchus congestus mirabilis} weevil. Scale bar: \SI{0.5}{\centi\meter} \cite{djeghdi2022}. (B) A section of the weevil's skeletonized photonic network, giving rise to blue structural coloration. 
    The rendered color corresponds to the simulated optical response for the experimental mean bond length $d=\SI{160}{\nano\meter}$ \cite{djeghdi2022}. (C) The biological reflectance spectrum compared to the average reflectance spectra for diamond-like networks and \textbf{ctn}-like networks, both at $\beta=0.001$. For all networks, we set the bond radius to $R=0.39$.
    (D) The average misfit functions comparing reflectance spectra of the biological network with three different light propagation directions to average computer-generated reflectances. (E) The critical pore radius $\delta_\mathrm{c}$ of the biological network and computer-generated networks. (F) The mean ring radius $\overline{r}_s$ of the biological network and computer-generated networks. (G) The ring radius standard deviation $\sigma_{r_s}$ of the biological network and computer-generated networks.}
    \label{fig:biological}
\end{figure*}

We test the applicability of the inverse method outside of computer-generated networks by investigating the color-structure relation of the biophotonic network creating the blue structural color of the \textit{Pachyrhynchus congestus mirabilis} weevil \cite{djeghdi2022} (Figures~\ref{fig:biological}A-B). Figure~\ref{fig:biological}C shows that diamond- and \textbf{ctn}-like networks at small $\beta$ closely reproduce the biological spectrum. The misfit function lacks a distinct minimum and yields similarly low values for these network types and $\beta$ values (Table~\ref{tab:biological_misfit}), whereas the minimum misfits to all other network types are more than twice as large. We attribute the poor agreement of primitive-cubic- and gyroid-like networks to their substantially different ring and pore sizes: the biological network is characterized by $\delta_\mathrm{c}=0.31$ and $\overline{r}_s=0.71$, values that lie far from those of these network types (Figure~\ref{fig:networks_order_metrics}E).

\begin{table}[tbp]
\centering
\caption{Misfit between biological and computer-generated spectra}
\label{tab:biological_misfit}
\begin{tabular}{lrrr}
Network type &
$R$ &
$\beta$ &
$\langle \chi(\beta) \rangle$ \\
\midrule
Primitive cubic & 0.39 & 0.01 & $0.00527\pm0.00026$ \\
Primitive cubic & 0.23 & 0.01 & $0.00175\pm0.00019$ \\
Diamond & 0.39 & 0.001 & $0.00029\pm0.00005$ \\
\textbf{ctn} & 0.39 & 0.001 & $0.00036\pm0.00005$ \\
Gyroid & 0.39 & 0.2 & $0.00087\pm0.00010$ \\
Gyroid & 0.54 & 0.2 & $0.00178\pm0.00018$ \\
\bottomrule
\end{tabular}

\iftoggle{isPNAS}
{\addtabletext{For each network type and bond radius $R$, the table lists the minimum average misfit between the biological reflectance spectrum and the computer-generated networks, together with the corresponding $\beta$ value. Within the uncertainty of the analysis, diamond- and \textbf{ctn}-like networks provide the closest match to the biological spectrum (Figure~\ref{fig:biological}C).}}
{\caption{For each network type and bond radius $R$, the table lists the minimum average misfit between the biological reflectance spectrum and the computer-generated networks, together with the corresponding $\beta$ value. Within the uncertainty of the analysis, diamond- and \textbf{ctn}-like networks provide the closest match to the biological spectrum (Figure~\ref{fig:biological}C).}}
\end{table}

We increase the $\beta$ resolution at small values for diamond- and \textbf{ctn}-like networks by extending the dataset of disordered networks, and plot the average misfit $\langle \chi (\beta) \rangle$ between the biological and the computer-generated spectra (Figure~\ref{fig:biological}D). The diamond-like misfit function is minimal in a broad region for $\beta \lesssim 0.2$, whereas the \textbf{ctn}-like misfit function shows a local minimum at $\beta=0.001$ and another minimal misfit of comparable depth at $\beta=0.2$, both slightly above the diamond-like plateau. To explain these features, we compare the pore and ring statistics of the biological and computer-generated networks (Figure~\ref{fig:biological}E-G).

The critical pore radius $\delta_\mathrm{c}$ of diamond-like networks closely matches the biological value over the entire $\beta$ range. In contrast, the $\delta_\mathrm{c}$ values of \textbf{ctn}-like networks are consistently larger and closely follow the shape of $\langle \chi(\beta)\rangle$, explaining the elevated misfits for intermediate $\beta$ values (Figure~\ref{fig:biological}D-E). For $\beta \gtrsim 0.2$, $\delta_\mathrm{c}$ remains approximately constant, while the misfit increases because both the mean ring radius $\overline{r}_s$ and its disorder $\sigma_{r_s}$ increasingly deviate from the biological values (Figure~\ref{fig:biological}F-G).

\begin{figure}[h]
\centering
\includegraphics[width=\iftoggle{isPNAS}
{0.48\linewidth}{0.66\linewidth}]{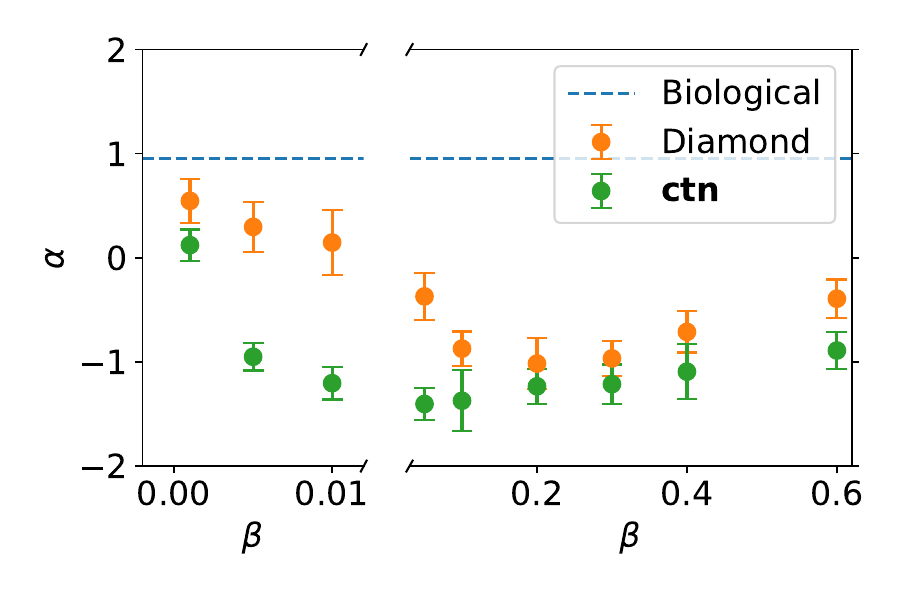}
\caption{\label{fig:hyperuniformity_bio_ctn_dia}
The hyperuniformity metric $\alpha$ of the biological network and computer-generated networks.}
\end{figure}

At small $\beta$, however, the mean ring radius of \textbf{ctn}-like networks agrees more closely with the biological value than does that of diamond-like networks, partially compensating for the mismatch in $\delta_\mathrm{c}$. Based on pore and ring statistics alone, one would therefore expect the optimal diamond-like networks to occur near $\beta=0.2$. We attribute the slightly lower misfit at $\beta=0.001$ to hyperuniformity. Whereas all networks with $\beta \geq 0.05$ are nonhyperuniform, the networks generated at $\beta=0.001$ exhibit the highest values of the hyperuniformity metric: $\alpha=0.55\pm0.21$ for diamond-like and $\alpha=0.12\pm0.15$ for \textbf{ctn}-like networks (\iftoggle{isPNAS}
{Figure~S1, Supporting Information}{Figure~\ref{fig:hyperuniformity_bio_ctn_dia}, Appendix}). The biological network is even more hyperuniform ($\alpha=0.95$).

Nevertheless, the misfit difference between the hyperuniform networks at $\beta=0.001$ and the nonhyperuniform networks at $\beta=0.2$ is small, suggesting that hyperuniformity only weakly affects the blue structural-color spectra. We likewise find no correlation between the misfit and other order metrics, including bond-length and bond-angle disorder ($\sigma_r$ and $\sigma_\theta$). Similarly, despite the biological and \textbf{ctn}-like networks sharing a comparable coordination statistic ($Z\approx3.4\pm0.5$), diamond-like networks ($Z=4.0$) reproduce the biological spectrum more accurately\footnote{The true biological coordination number may be higher than the reported value of $Z\approx3.4\pm0.5$, because binarization and skeletonization tend to underestimate $Z$ \cite{djeghdi2022}. This uncertainty does not affect our argument because the computer-generated spectra are compared with the experimental data and its reported coordination number.}. This agreement is unexpected given that primitive similarity, alongside hyperuniformity, has previously been identified as a key factor governing the optical response of disordered photonic materials. \cite{florescu2009, sellers2017}. In contrast, our results indicate that these characteristics play only a minor role in the blue structural-color response. Instead, the spectra are predominantly governed by pore and ring geometry, supporting our hypothesis that these motifs act as the relevant local resonators.

A high degree of ring similarity may enhance structural-color saturation through multiple mechanisms. Yang et al.~\cite{yang2021} showed that uniform ring geometries facilitate the localization of electromagnetic energy within a single material phase, a key ingredient for PBG formation at high dielectric contrast. We further suggest coherent backscattering as a complementary mechanism \cite{akkermans1986}. Because of time-reversal symmetry, counterpropagating light paths around a network ring interfere, giving rise to weak localization and enhanced reflectance \cite{lagendijk2009}. As ring similarity increases, the associated resonance frequencies become more narrowly distributed, producing a sharper reflectance peak and, consequently, more saturated structural color.

Our finding that pores act as the relevant resonators in photonic networks, with their interference mediated by short-range spatial correlations, naturally suggests an analogy to an inverse photonic glass. Like direct and inverse photonic glasses, all networks considered here exhibit high reflectance above a threshold frequency, favoring blue and green structural colors while suppressing saturated reds. In photonic glass materials, this characteristic response arises from the interplay of Mie and cavity resonances with multiple scattering \cite{magkiriadou2014, shang2020, jacucci2020}. Consistent with this picture, most reported biological disordered photonic networks produce blue or green coloration \cite{sellers2017, djeghdi2022, bauernfeind2023, bauernfeind2024}. The only known exception, the red-to-orange network of the longhorn beetle \textit{Sternotomis amabilis}, generates a comparatively desaturated brownish color \cite{bauernfeind2024}.

Approximating the pores of disordered photonic networks as monodisperse spheres is clearly a simplification. Nevertheless, Jacucci et al.~\cite{jacucci2020} showed that the structural color of inverse photonic glasses is remarkably robust to polydispersity, considerably more so than in direct photonic glasses. They further found that inverse systems produce more saturated colors because structural correlations dominate over form-factor contributions, which otherwise increase the high-frequency reflectance and reduce color saturation.
Equation~\ref{eqn:volume_fraction_pbg} suggests an additional explanation for the enhanced color response of inverse photonic glasses. Owing to their lower volume fraction, the electromagnetic field can be more strongly confined to a single phase near the color peak, a condition that would broaden the PBG at higher dielectric contrast. At the lower dielectric contrasts characteristic of biological materials, the residual effect of this PBG may still reduce the photonic density of states, thereby enhancing the color peak.

These insights demonstrate that, unlike photonic networks, the spherical building blocks of direct and inverse photonic glasses permit a more direct theoretical description. The extensive understanding of these systems, therefore, provides a useful framework for interpreting disordered photonic networks and may inspire new strategies for designing structural-color materials.

\section*{Conclusion}

Structural color arises from disordered biophotonic networks, even at a relatively low dielectric contrast. The statistical complexity of disordered nanostructures, combined with the high experimental and computational costs of their analysis, has long hindered a mechanistic understanding of their structural color. We used an inverse method to investigate which structural information is encoded in the simulated reflectance spectra of computer-generated and biological networks.

Our analysis revealed that all bicontinous systems exhibit low reflectance at low frequencies in the homogenized regime and a reflectance peak as the frequency enters the resonant regime. We identified network rings and pores as the relevant local scatterers and demonstrated that increased ring disorder broadens the reflectance peak. Ring and pore sizes anticorrelate with the mean coordination number that characterizes the network type, so that a decrease in coordination induces larger resonators and, thus, a red-shifted spectrum.
Both spectral signatures are captured by a misfit function that compares a target spectrum with the average spectra of the system. We inversely obtain relevant information on the network type and degree of disorder from the location of the misfit-function minimum.

We used the inverse method's misfit function to show that the blue color of the disordered biophotonic network in the \textit{Pachyrhynchus congestus mirabilis} weevil is characterized by its ring and pore size statistics. On the other hand, we found no significant effect of hyperuniformity and primitive similarity, suggesting that biophotonic blue structural color is not a remnant of a PBG at higher dielectric contrast. Instead, we proposed an analogy between the network pores and the spherical inclusions in inverse photonic glasses \cite{garcia2007, jacucci2020}. In these systems, the high reflectance above a threshold frequency is understood in terms of the interplay between local resonances and multiple scattering \cite{magkiriadou2014, shang2020, jacucci2020}. We attributed the predominance of blue-to-green biophotonic networks to the correspondence between photonic networks and glasses.

The design and fabrication of bio-inspired photonic materials is an active field of research \cite{datta2022, vogler-neuling2023, demirors2024}. Due to their brilliance, durability, and bio-compatibility, these systems have applications in paints, dyes, camouflage, sensing, and displays \cite{dong2022, arora2024}.
Among structural color materials, 3D disordered photonic networks are particularly attractive because they combine angle-independent optical responses with a vast design space and numerous structural degrees of freedom. Yet this very complexity has hindered the development of predictive design rules and the targeted fabrication of such materials.
Our work reveals which geometric motifs affect the structural color spectrum in biological photonic networks. These insights establish new principles for the design and targeted fabrication of bio-inspired structural color materials.

\vspace{2\baselineskip}
\noindent
{\Large \textbf{Methods}}

\iftoggle{isPNAS}{\matmethods}{}{

\subsection*{Computer-Generated Disordered Networks}

Disordered networks were generated using an extended Wooten-Weaire-Winer algorithm \cite{wooten1985, hemmann2026}. Starting from a periodic crystalline network, disorder was introduced through bond-switch moves. Moves were accepted with the Metropolis probability
$P_\mathrm{accept}=\min \left[1, \exp \left( \frac{E_\mathrm{f}-E_\mathrm{i}}{T} \right) \right]$,
where $T$ is a temperature parameter and $E_\mathrm{i}$ and $E_\mathrm{f}$ are the initial and final strain energies \cite{metropolis1953}. The strain energy was described by a modified Keating potential \cite{keating1966, hemmann2026}
\begin{equation*}
    E=
    \frac{3}{16}
    \Sigma_{\langle ij \rangle}
    \left(
    r_{ij}^2-1
    \right)^2
    +
    \frac{3}{8}
    \beta
    \Sigma_{\langle jik \rangle}
    \left(
    r_{ij} r_{ik} \cos(\theta_{jik}) + 1
    \right)^2
    \quad ,
\end{equation*}
where $r_{ij}$ is the bond length (in units of the equilibrium bond length $d$), $\theta_{jik}$ is the bond angle, and $\beta \in [0.01,10.0]$ is the bond-bending force constant. A triangular temperature profile was employed, consisting of heating at a rate $\Delta T$ to $T_\mathrm{max}$, cooling at $-\Delta T$, and a final quench at $T=0$ \cite{hemmann2026}. We generated disordered networks starting from primitive cubic, diamond, \textbf{ctn}, and gyroid networks with cubic supercells containing 512 vertices (primitive cubic, diamond, and gyroid) or 728 vertices (\textbf{ctn}) with edge lengths $8.0d$, $9.24d$, $11.11d$, and $11.31d$, respectively. Periodic boundary conditions were applied in all directions. The temperature profile parameters were $T_\mathrm{max}=1.5\,T_\mathrm{melt}$ for primitive-cubic, diamond, and \textbf{ctn} networks, $T_\mathrm{max}=2.0\,T_\mathrm{melt}$ for gyroid networks, and $\Delta T=0.5\,T_\mathrm{melt}$ for all systems. For each network type, 25 realizations were generated for each of 13 $\beta$ values, yielding 325 configurations in total, of which 195 were used for averaging and 130 for testing.

\subsection*{Finite-Difference Time-Domain Simulations}

For each generated network, we ran three finite-difference time-domain simulations with light propagating along (100), (010), and (001) orientations to obtain reflectance spectra. Network bonds were decorated with cylinders with refractive index $n_\mathrm{m}=1.5$, close to that of chitin, which biophotonic networks in beetles and weevils commonly consist of \cite{djeghdi2022, bauernfeind2023, bauernfeind2024}. The background is vacuum $n_\mathrm{b}=1$.
Unless otherwise stated, the cylinder radius, in units of $d$, is set to $R=0.39$, corresponding to a volume fraction $\phi=0.4$ in the biophotonic network of \textit{Pachyrhynchus congestus mirabilis} \cite{djeghdi2022} and close to the optimum predicted by Equation~\eqref{eqn:volume_fraction_pbg} for $n_\mathrm{m}=1.5$ and $n_\mathrm{b}=1$. An electrical conductivity of $\sigma=\SI{1e-4}{\siemens}/d$ was introduced to suppress high-order Fabry-Pérot resonances. The simulation domain was periodic in the directions orthogonal to light propagation and terminated by absorbing boundaries along the propagation direction. The lateral dimensions were equal to the supercell edge length, and the slab thickness was fixed at $9.0d$. For primitive cubic networks, whose supercell edge length is smaller than the slab thickness, the network repeated once along the propagation direction. For each initial network type, 585 spectra (45 per $\beta$ value) were used for averaging and 390 spectra (30 per $\beta$ value) for testing.

Because the biological network is finite and nonperiodic, the simulation cell boundaries, orthogonal to the light-propagation direction, contain dangling bonds and truncated cylinders. To reduce boundary effects, the simulation box dimensions were fixed to $24d\times24d$ in the $x$ and $y$ directions, slightly below the minimum lateral size of the cuboidal experimental networks. The slab thicknesses, bond radii, and simulated orientations [(100), (010), and (001)] were chosen identically to those of the computer-generated networks.

\subsection*{AI-assisted writing}
Microsoft Copilot and Grammarly were used during manuscript preparation to assist with text drafting and editing. The tool was not used for data analysis, interpretation of results, figure generation, or scientific decision-making. All scientific content was reviewed and validated by the authors.

}

\vspace{2\baselineskip}
\noindent
{\Large \textbf{Data Availability}}

\noindent
The data supporting the findings of this article are available upon request from the authors.

\vspace{2\baselineskip}
\noindent
{\Large \textbf{Acknowledgments}}

\noindent
This study was supported by a European Research Council (ERC) Advanced grant (PrISMoID, 833895), the Adolphe Merkle Foundation, and the Swiss National Science Foundation (SNSF) through the National Center of Competence in Research Bio-Inspired Materials (grant no. 51NF40-182881). F.A. P. acknowledges financial support from CNPq, CAPES, and FAPERJ. This publication has emanated from research conducted with the financial support of Research Ireland, Grant Number 12/RC/2278\_2, and is co-funded under the European Regional Development Fund under the AMBER award.

\vspace{2\baselineskip}
\noindent
{\Large \textbf{Appendix: Correlating spectral and structural misfit }}
\iftoggle{isPNAS}
{\section*{Correlating spectral and structural misfit}}{}

\FloatBarrier

\FloatBarrier

\balance
\bibliographystyle{MSP}

\bibliography{refs}

@article{akkermans1986,
	title = {Coherent Backscattering of Light by Disordered Media: Analysis of the Peak Line Shape},
	volume = {56},
	url = {https://link.aps.org/doi/10.1103/PhysRevLett.56.1471},
	doi = {10.1103/PhysRevLett.56.1471},
	shorttitle = {Coherent Backscattering of Light by Disordered Media},
	pages = {1471--1474},
	number = {14},
	journal = {Physical Review Letters},
	shortjournal = {Phys. Rev. Lett.},
	publisher = {American Physical Society},
	author = {Akkermans, E. and Wolf, P. E. and Maynard, R.},
	urldate = {2026-07-31},
	date = {1986-04-07},
	year = {1986},
}

@article{arora2024,
  author  = {Arora, Ekta Kundra and Sharma, Vibha and Sethi, Geetanjali and Puthanagady, Mariet Sibi and Meena, Anjali},
  title   = {Bioinspired Designer Surface Nanostructures for Structural Color},
  journal = {Nanotechnology for Environmental Engineering},
  year    = {2024},
  volume  = {9},
  number  = {3},
  pages   = {461--472},
  doi     = {10.1007/s41204-024-00368-7},
  url     = {https://doi.org/10.1007/s41204-024-00368-7}
}

@article{bauernfeind2023,
	title = {Photonic Amorphous {I-WP}-Like Networks Create Angle-Independent Colors in Sternotomis virescens Longhorn Beetles},
	volume = {34},
	issn = {1616-3028},
	url = {https://onlinelibrary.wiley.com/doi/abs/10.1002/adfm.202302720},
	doi = {10.1002/adfm.202302720},
	pages = {2302720},
	issue = {n/a},
	journal = {Advanced Functional Materials},
	author = {Bauernfeind, Viola and Djeghdi, Kenza and Gunkel, Ilja and Steiner, Ullrich and Wilts, Bodo D.},
	urldate = {2023-05-25},
	date = {2023-05-18},
    year = {2023},
	langid = {english},
}

@article{bauernfeind2024,
	title = {Not only a matter of disorder in I-{WP} minimal surface-based photonic networks: Diffusive structural color in \textit{Sternotomis amabilis} longhorn beetles},
	volume = {23},
	issn = {2590-0498},
	url = {https://www.sciencedirect.com/science/article/pii/S2590049824000614},
	doi = {10.1016/j.mtadv.2024.100524},
	journal = {Materials Today Advances},
	shortjournal = {Materials Today Advances},
	author = {Bauernfeind, Viola and Saranathan, Vinodkumar and Djeghdi, Kenza and Longo, Elena and Flenner, Silja and Greving, Imke and Steiner, Ullrich and Wilts, Bodo D.},
	urldate = {2025-09-09},
	date = {2024-08-01},
    year = {2024},
}

@article{datta2022,
author = {Datta, Bianca and Spero, Ellan F. and Martin-Martinez, Francisco J. and Ortiz, Christine},
title = {Socially-Directed Development of Materials for Structural Color},
journal = {Advanced Materials},
volume = {34},
number = {20},
pages = {2100939},
doi = {https://doi.org/10.1002/adma.202100939},
url = {https://advanced.onlinelibrary.wiley.com/doi/abs/10.1002/adma.202100939},
eprint = {https://advanced.onlinelibrary.wiley.com/doi/pdf/10.1002/adma.202100939},
year = {2022}
}

@article{djeghdi2022,
	title = {{3D} Tomographic Analysis of the Order-Disorder Interplay in the \textit{Pachyrhynchus congestus mirabilis} Weevil},
	volume = {9},
	issn = {2198-3844},
	url = {https://onlinelibrary.wiley.com/doi/abs/10.1002/advs.202202145},
	doi = {10.1002/advs.202202145},
	pages = {2202145},
	number = {26},
	journal = {Advanced Science},
	author = {Djeghdi, Kenza and Steiner, Ullrich and Wilts, Bodo D.},
	urldate = {2023-05-25},
	year = {2022},
	langid = {english},
}

@article{demirors2024,
	title = {Tuning disorder in structurally colored bioinspired photonic glasses},
	volume = {20},
	issn = {1744-683X},
	url = {https://doi.org/10.1039/d3sm01468a},
	doi = {10.1039/d3sm01468a},
	pages = {1620--1628},
	number = {7},
	journal = {Soft Matter},
	shortjournal = {Soft Matter},
	author = {Demirörs, Ahmet F. and Manne, Kalpana and Magkiriadou, Sofia and Scheffold, Frank},
	urldate = {2026-07-24},
	date = {2024-02-21},
	year = {2024},
}

@article{dong2022,
title = {Bio-inspired non-iridescent structural coloration enabled by self-assembled cellulose nanocrystal composite films with balanced ordered/disordered arrays},
journal = {Composites Part B: Engineering},
volume = {229},
pages = {109456},
year = {2022},
issn = {1359-8368},
doi = {https://doi.org/10.1016/j.compositesb.2021.109456},
url = {https://www.sciencedirect.com/science/article/pii/S1359836821008234},
author = {Xiu Dong and Ze-Lian Zhang and Yu-Yao Zhao and Dong Li and Zi-Li Wang and Chen Wang and Fei Song and Xiu-Li Wang and Yu-Zhong Wang},
}

@article{economou1993,
	title = {Classical wave propagation in periodic structures: Cermet versus network topology},
	volume = {48},
	issn = {0163-1829, 1095-3795},
	url = {https://link.aps.org/doi/10.1103/PhysRevB.48.13434},
	doi = {10.1103/PhysRevB.48.13434},
	shorttitle = {Classical wave propagation in periodic structures},
	pages = {13434--13438},
	number = {18},
	journal = {Physical Review B},
	shortjournal = {Phys. Rev. B},
	author = {Economou, E. N. and Sigalas, M. M.},
	urldate = {2023-05-25},
	date = {1993-11-01},
	year = {1993},
	langid = {english},
}

@article{edagawa2008,
	title = {Photonic Amorphous Diamond Structure with a {3D} Photonic Band Gap},
	volume = {100},
	issn = {0031-9007, 1079-7114},
	url = {https://link.aps.org/doi/10.1103/PhysRevLett.100.013901},
	doi = {10.1103/PhysRevLett.100.013901},
	pages = {013901},
	number = {1},
	journal = {Physical Review Letters},
	shortjournal = {Phys. Rev. Lett.},
	author = {Edagawa, Keiichi and Kanoko, Satoshi and Notomi, Masaya},
	urldate = {2023-06-23},
	date = {2008-01-02},
    year = {2008},
	langid = {english},
}

@article{florescu2009,
	title = {Designer disordered materials with large, complete photonic band gaps},
	volume = {106},
	issn = {0027-8424, 1091-6490},
	url = {https://pnas.org/doi/full/10.1073/pnas.0907744106},
	doi = {10.1073/pnas.0907744106},
	pages = {20658--20663},
	number = {49},
	journal = {Proceedings of the National Academy of Sciences},
	shortjournal = {Proc. Natl. Acad. Sci. U.S.A.},
	author = {Florescu, Marian and Torquato, Salvatore and Steinhardt, Paul J.},
	urldate = {2023-05-25},
	date = {2009-12-08},
    year = {2009},
	langid = {english},
}

@article{garcia2007,
  author       = {Garc\'{\i}a, P. D. and Sapienza, R. and Blanco, \'{A}. and L\'opez, C.},
  title        = {Photonic Glass: A Novel Random Material for Light},
  journal      = {Advanced Materials},
  year         = {2007},
  volume       = {19},
  number       = {18},
  pages        = {2597--2602},
  doi          = {10.1002/adma.200602426},
  url          = {https://onlinelibrary.wiley.com/doi/abs/10.1002/adma.200602426}
}

@article{hemmann2026,
author = {Hemmann, Florin and Glauser, Vincent and Steiner, Ullrich and Saba, Matthias},
title = {Algorithmic Design of Disordered Networks With Arbitrary Coordination: Application to Biophotonics},
journal = {Advanced Functional Materials},
volume = {36},
number = {38},
pages = {e00037},
doi = {https://doi.org/10.1002/adfm.202600037},
url = {https://advanced.onlinelibrary.wiley.com/doi/abs/10.1002/adfm.202600037},
eprint = {https://advanced.onlinelibrary.wiley.com/doi/pdf/10.1002/adfm.202600037},
year = {2026}
}

@article{jacucci2020,
	title = {The limitations of extending nature’s color palette in correlated, disordered systems},
	volume = {117},
	url = {https://www.pnas.org/doi/10.1073/pnas.2010486117},
	doi = {10.1073/pnas.2010486117},
	pages = {23345--23349},
	number = {38},
	journal = {Proceedings of the National Academy of Sciences},
	publisher = {Proceedings of the National Academy of Sciences},
	author = {Jacucci, Gianni and Vignolini, Silvia and Schertel, Lukas},
	urldate = {2026-07-27},
	date = {2020-09-22},
	year = {2020},
}

@book{joannopoulos2008,
 ISBN = {9780691124568},
 URL = {http://www.jstor.org/stable/j.ctvcm4gz9},
 author = {John D. Joannopoulos and Steven G. Johnson and Joshua N. Winn and Robert D. Meade},
 edition = {Second Edition},
 publisher = {Princeton University Press},
 title = {Photonic Crystals: Molding the Flow of Light},
 urldate = {2025-11-18},
 year = {2008}
}

@article{keating1966,
	title = {Effect of Invariance Requirements on the Elastic Strain Energy of Crystals with Application to the Diamond Structure},
	volume = {145},
	url = {https://link.aps.org/doi/10.1103/PhysRev.145.637},
	doi = {10.1103/PhysRev.145.637},
	pages = {637--645},
	number = {2},
	journal = {Physical Review},
	shortjournal = {Phys. Rev.},
	author = {Keating, P. N.},
	urldate = {2023-10-12},
	date = {1966-05-13},
    year = {1966},
}

@article{klatt2022,
	title = {Wave propagation and band tails of two-dimensional disordered systems in the thermodynamic limit},
	volume = {119},
	url = {https://www.pnas.org/doi/10.1073/pnas.2213633119},
	doi = {10.1073/pnas.2213633119},
	pages = {e2213633119},
	number = {52},
	journal = {Proceedings of the National Academy of Sciences},
	publisher = {Proceedings of the National Academy of Sciences},
	author = {Klatt, Michael A. and Steinhardt, Paul J. and Torquato, Salvatore},
	urldate = {2024-03-28},
	date = {2022-12-27},
	year = {2022},
}

@article{lagendijk2009,
author = {Lagendijk, Ad and Tiggelen, Bart and Wiersma, Diederik},
year = {2009},
month = {08},
pages = {24-29},
title = {Fifty years of Anderson localization},
volume = {62},
journal = {Physics Today},
doi = {10.1063/1.3206091}
}

@article{liew2011,
	title = {Photonic band gaps in three-dimensional network structures with short-range order},
	volume = {84},
	url = {https://link.aps.org/doi/10.1103/PhysRevA.84.063818},
	doi = {10.1103/PhysRevA.84.063818},
	pages = {063818},
	number = {6},
	journal = {Physical Review A},
	shortjournal = {Phys. Rev. A},
	publisher = {American Physical Society},
	author = {Liew, Seng Fatt and Yang, Jin-Kyu and Noh, Heeso and Schreck, Carl F. and Dufresne, Eric R. and O’Hern, Corey S. and Cao, Hui},
	year = {2011},
	urldate = {2023-07-19},
	date = {2011-12-07},
}

@article{macedo2026,
	title = {Inverse determination of light-matter coupling in disordered systems from transmittance spectra},
	volume = {113},
	url = {https://link.aps.org/doi/10.1103/cggf-fpyb},
	doi = {10.1103/cggf-fpyb},
	pages = {214202},
	number = {21},
	journal = {Physical Review B},
	shortjournal = {Phys. Rev. B},
	publisher = {American Physical Society},
	author = {Macedo, Thales F. and Faúndez, Julián and Coelho, Antônio S. and Lewenkopf, Caio and Ferreira, Mauro S. and Pinheiro, Felipe A. and Costa, Natanael C.},
	urldate = {2026-07-23},
	date = {2026-06-08},
	year = {2026},
}

@article{magkiriadou2014,
	title = {Absence of red structural color in photonic glasses, bird feathers, and certain beetles},
	volume = {90},
	url = {https://link.aps.org/doi/10.1103/PhysRevE.90.062302},
	doi = {10.1103/PhysRevE.90.062302},
	pages = {062302},
	number = {6},
	journal = {Physical Review E},
	shortjournal = {Phys. Rev. E},
	publisher = {American Physical Society},
	author = {Magkiriadou, Sofia and Park, Jin-Gyu and Kim, Young-Seok and Manoharan, Vinothan N.},
	urldate = {2026-07-27},
	date = {2014-12-03},
	year = {2014},
}

@article{metropolis1953,
	title = {Equation of State Calculations by Fast Computing Machines},
	volume = {21},
	issn = {0021-9606},
	url = {https://doi.org/10.1063/1.1699114},
	doi = {10.1063/1.1699114},
	pages = {1087--1092},
	number = {6},
	journal = {The Journal of Chemical Physics},
	shortjournal = {The Journal of Chemical Physics},
	author = {Metropolis, Nicholas and Rosenbluth, Arianna W. and Rosenbluth, Marshall N. and Teller, Augusta H. and Teller, Edward},
	urldate = {2024-02-26},
	date = {1953-06-01},
    year = {1953},
}

@article{mukim2020,
	title = {Disorder information from conductance: A quantum inverse problem},
	volume = {102},
	url = {https://link.aps.org/doi/10.1103/PhysRevB.102.075409},
	doi = {10.1103/PhysRevB.102.075409},
	shorttitle = {Disorder information from conductance},
	pages = {075409},
	number = {7},
	journal = {Physical Review B},
	shortjournal = {Phys. Rev. B},
	publisher = {American Physical Society},
	author = {Mukim, S. and Amorim, F. P. and Rocha, A. R. and Muniz, R. B. and Lewenkopf, C. and Ferreira, M. S.},
	urldate = {2026-03-02},
	date = {2020-08-05},
	year = {2020},
}

@article{mukim2022,
title = {Spatial mapping of disordered {2D} systems: The conductance {Sudoku}},
journal = {Carbon},
volume = {188},
pages = {360-366},
year = {2022},
issn = {0008-6223},
doi = {https://doi.org/10.1016/j.carbon.2021.11.073},
url = {https://www.sciencedirect.com/science/article/pii/S0008622321011593},
author = {S. Mukim and C. Lewenkopf and M.S. Ferreira},
}

@misc{rcsr_ctn2025,
  author = {Olaf Delgado-Friedrichs and Michael O'Keeffe},
  title = {ctn},
  year = 2025,
  url = {http://rcsr.net/nets/ctn},
  urldate = {2025-11-17}
}

@article{rothammer2021,
	title = {Tailored Disorder in Photonics: Learning from Nature},
	volume = {9},
	issn = {2195-1071, 2195-1071},
	url = {https://onlinelibrary.wiley.com/doi/10.1002/adom.202100787},
	doi = {10.1002/adom.202100787},
	shorttitle = {Tailored Disorder in Photonics},
	pages = {2100787},
	number = {19},
	journal = {Advanced Optical Materials},
	shortjournal = {Advanced Optical Materials},
	author = {Rothammer, Maximilian and Zollfrank, Cordt and Busch, Kurt and Von Freymann, Georg},
	urldate = {2023-05-25},
	date = {2021-10},
    year = {2021},
	langid = {english},
}

@article{schertel2019,
	title = {The Structural Colors of Photonic Glasses},
	volume = {7},
	rights = {© 2019 {WILEY}-{VCH} Verlag {GmbH} \& Co. {KGaA}, Weinheim},
	issn = {2195-1071},
	url = {https://onlinelibrary.wiley.com/doi/abs/10.1002/adom.201900442},
	doi = {10.1002/adom.201900442},
	pages = {1900442},
	number = {15},
	journal = {Advanced Optical Materials},
	author = {Schertel, Lukas and Siedentop, Lukas and Meijer, Janne-Mieke and Keim, Peter and Aegerter, Christof M. and Aubry, Geoffroy J. and Maret, Georg},
	urldate = {2026-07-27},
	date = {2019},
	year = {2019},
	langid = {english},
}

@article{sellers2017,
	title = {Local self-uniformity in photonic networks},
	volume = {8},
	issn = {2041-1723},
	url = {https://www.nature.com/articles/ncomms14439},
	doi = {10.1038/ncomms14439},
	pages = {14439},
	number = {1},
	journal = {Nature Communications},
	shortjournal = {Nat Commun},
	author = {Sellers, Steven R. and Man, Weining and Sahba, Shervin and Florescu, Marian},
	urldate = {2023-05-25},
	date = {2017-02-17},
    year = {2017},
	langid = {english},
}

@article{siedentop2024,
	title = {Stealthy and hyperuniform isotropic photonic band gap structure in {3D}},
	volume = {3},
	issn = {2752-6542},
	url = {https://doi.org/10.1093/pnasnexus/pgae383},
	doi = {10.1093/pnasnexus/pgae383},
	pages = {383},
	number = {9},
	journal = {{PNAS} Nexus},
	shortjournal = {{PNAS} Nexus},
	author = {Siedentop, Lukas and Lui, Gianluc and Maret, Georg and Chaikin, Paul M and Steinhardt, Paul J and Torquato, Salvatore and Keim, Peter and Florescu, Marian},
	urldate = {2024-10-22},
    year = {2024},
	date = {2024-09-01},
}

@article{shang2020,
	title = {Photonic glass based structural color},
	volume = {5},
	issn = {2378-0967},
	url = {https://doi.org/10.1063/5.0006203},
	doi = {10.1063/5.0006203},
	pages = {060901},
	number = {6},
	journal = {{APL} Photonics},
	shortjournal = {{APL} Photonics},
	author = {Shang, Guoliang and Eich, Manfred and Petrov, Alexander},
	urldate = {2026-07-27},
	date = {2020-06-08},
	year = {2020},
}

@article{torquato2018,
title = {Hyperuniform states of matter},
journal = {Physics Reports},
volume = {745},
pages = {1-95},
year = {2018},
issn = {0370-1573},
doi = {https://doi.org/10.1016/j.physrep.2018.03.001},
url = {https://www.sciencedirect.com/science/article/pii/S037015731830036X},
author = {Salvatore Torquato}
}

@article{vogler-neuling2023,
  title = {Biopolymer {{Photonics}}: {{From Nature}} to {{Nanotechnology}}},
  shorttitle = {Biopolymer {{Photonics}}},
  author = {Vogler-Neuling, Viola V. and Saba, Matthias and Gunkel, Ilja and Zoppe, Justin O. and Steiner, Ullrich and Wilts, Bodo D. and Dodero, Andrea},
  year = {2023},
  month = aug,
  journal = {Advanced Functional Materials},
  volume = {34},
  pages = {2306528},
  issn = {1616-301X, 1616-3028},
  doi = {10.1002/adfm.202306528},
  url = {https://doi.org/10.1002/adfm.202306528}
}

@article{vukusic2003,
	title = {Photonic structures in biology},
	volume = {424},
	rights = {2003 Springer Nature Limited},
	issn = {1476-4687},
	url = {https://www.nature.com/articles/nature01941},
	doi = {10.1038/nature01941},
	pages = {852--855},
	number = {6950},
	journal = {Nature},
	publisher = {Nature Publishing Group},
	author = {Vukusic, Pete and Sambles, J. Roy},
	urldate = {2026-07-27},
	date = {2003-08},
	year = {2003},
	langid = {english},
}

@article{wooten1985,
	title = {Computer Generation of Structural Models of Amorphous {Si} and {Ge}},
	volume = {54},
	url = {https://link.aps.org/doi/10.1103/PhysRevLett.54.1392},
	doi = {10.1103/PhysRevLett.54.1392},
	pages = {1392--1395},
	number = {13},
	journal = {Physical Review Letters},
	shortjournal = {Phys. Rev. Lett.},
	author = {Wooten, F. and Winer, K. and Weaire, D.},
	urldate = {2023-06-23},
	date = {1985-04-01},
    year = {1985},
}

@article{yang2021,
	title = {Ring uniformity in amorphous photonic band gap materials},
	volume = {104},
	url = {https://link.aps.org/doi/10.1103/PhysRevB.104.054208},
	doi = {10.1103/PhysRevB.104.054208},
	pages = {054208},
	number = {5},
	journal = {Physical Review B},
	shortjournal = {Phys. Rev. B},
	publisher = {American Physical Society},
	author = {Yang, Chih-Ying and Lai, Bo-Lin and Xie, Zhi-Hong and Hung, Yu-Chueh},
	urldate = {2025-01-06},
	date = {2021-08-24},
    year = {2021}
}

@book{yeh2005,
  title={Optical Waves in Layered Media},
  author={Yeh, P.},
  isbn={9780471731924},
  lccn={2004065926},
  series={Wiley Series in Pure and Applied Optics},
  url={https://books.google.ch/books?id=-yZBAQAAIAAJ},
  year={2005},
  publisher={Wiley}
}

@article{yin2012,
author = {Haiwei Yin  and Biqin Dong  and Xiaohan Liu  and Tianrong Zhan  and Lei Shi  and Jian Zi  and Eli Yablonovitch },
title = {Amorphous diamond-structured photonic crystal in the feather barbs of the scarlet macaw},
journal = {Proceedings of the National Academy of Sciences},
volume = {109},
number = {27},
pages = {10798-10801},
year = {2012},
doi = {10.1073/pnas.1204383109},
URL = {https://www.pnas.org/doi/abs/10.1073/pnas.1204383109},
eprint = {https://www.pnas.org/doi/pdf/10.1073/pnas.1204383109},}

@article{costa2026inferring,
  title={Inferring stealthy hyperuniform correlations from quantum transport},
  author={Costa, Natanael C and Ferreira, Mauro S and Lewenkopf, Caio and Pinheiro, Felipe A and Steinhardt, Paul J and Torquato, Salvatore and Vanoni, Carlo},
  journal={arXiv preprint arXiv:2608.11188},
  year={2026}
}

@article{vanoni2026effective,
  title={Effective delocalization in the one-dimensional Anderson model with stealthy disorder},
  author={Vanoni, Carlo and Karcher, Jonas and Rechtsman, Mikael C and Altshuler, Boris L and Steinhardt, Paul J and Torquato, Salvatore},
  journal={Physical review letters},
  volume={136},
  number={15},
  pages={150404},
  year={2026},
  publisher={APS}
}

\end{document}